\documentclass{mn2e}
\usepackage{graphicx}
\usepackage{color}

\begin{document}

\title[2dFLenS survey paper]{The 2-degree Field Lensing Survey: design and clustering measurements}

\author[Blake et al.]{\parbox[t]{\textwidth}{Chris
    Blake$^1$\footnotemark, Alexandra Amon$^2$, Michael
    Childress$^{3,4}$, Thomas Erben$^5$, \\ Karl Glazebrook$^1$,
    Joachim Harnois-Deraps$^{2,6}$, Catherine Heymans$^2$, \\ Hendrik
    Hildebrandt$^5$, Samuel R.\ Hinton$^7$, Steven Janssens$^8$,
    Andrew Johnson$^1$, \\ Shahab Joudaki$^1$, Dominik Klaes$^5$,
    Konrad Kuijken$^9$, Chris Lidman$^{10}$, \\ Felipe A.\ Marin$^1$,
    David Parkinson$^7$, Gregory B.\ Poole$^{11}$, Christian Wolf$^3$}
  \\ \\ $^1$ Centre for Astrophysics \& Supercomputing, Swinburne
  University of Technology, P.O.\ Box 218, Hawthorn, VIC 3122,
  Australia \\ $^2$ Scottish Universities Physics Alliance, Institute
  for Astronomy, University of Edinburgh, Royal Observatory, Blackford
  Hill, \\ Edinburgh, EH9 3HJ, U.K. \\ $^3$ Research School of
  Astronomy and Astrophysics, The Australian National University,
  Canberra, ACT 2611, Australia \\ $^4$ School of Physics and
  Astronomy, University of Southampton, Southampton, SO17 1BJ,
  U.K. \\ $^5$ Argelander Institute for Astronomy, University of Bonn,
  Auf dem Hugel 71, 53121, Bonn, Germany \\ $^6$ Department of Physics
  and Astronomy, University of British Columbia, 6224 Agricultural
  Road, Vancouver, V6T 1Z1, B.C., Canada \\ $^7$ School of Mathematics
  and Physics, University of Queensland, QLD 4072, Australia \\ $^8$
  Department of Astronomy, University of Toronto, Toronto, ON, M5S
  3H4, Canada \\ $^9$ Leiden Observatory, Leiden University, Niels
  Bohrweg 2, 2333 CA Leiden, The Netherlands \\ $^{10}$ Australian
  Astronomical Observatory, North Ryde, NSW 2113, Australia \\ $^{11}$
  School of Physics, University of Melbourne, Parkville, VIC 3010,
  Australia}

\maketitle

\begin{abstract}
We present the 2-degree Field Lensing Survey (2dFLenS), a new galaxy
redshift survey performed at the Anglo-Australian Telescope.  2dFLenS
is the first wide-area spectroscopic survey specifically targeting the
area mapped by deep-imaging gravitational lensing fields, in this case
the Kilo-Degree Survey.  2dFLenS obtained $70{,}079$ redshifts in the
range $z < 0.9$ over an area of $731$ deg$^2$, and is designed to
extend the datasets available for testing gravitational physics and
promote the development of relevant algorithms for joint imaging and
spectroscopic analysis.  The redshift sample consists first of
$40{,}531$ Luminous Red Galaxies (LRGs), which enable analyses of
galaxy-galaxy lensing, redshift-space distortion, and the overlapping
source redshift distribution by cross-correlation.  An additional
$28{,}269$ redshifts form a magnitude-limited ($r < 19.5$)
nearly-complete sub-sample, allowing direct source classification and
photometric-redshift calibration.  In this paper, we describe the
motivation, target selection, spectroscopic observations, and
clustering analysis of 2dFLenS.  We use power spectrum multipole
measurements to fit the redshift-space distortion parameter of the LRG
sample in two redshift ranges $0.15 < z < 0.43$ and $0.43 < z < 0.7$
as $\beta = 0.49 \pm 0.15$ and $\beta = 0.26 \pm 0.09$, respectively.
These values are consistent with those obtained from LRGs in the
Baryon Oscillation Spectroscopic Survey.  2dFLenS data products will
be released via our website {\tt http://2dflens.swin.edu.au}.
\end{abstract}

\section{Introduction}
\setcounter{footnote}{1}
\footnotetext{E-mail: cblake@astro.swin.edu.au}

A wide set of cosmological observations suggests that the dynamics of
the Universe are currently dominated by some form of `dark energy',
which in standard Friedmann-Robertson-Walker (FRW) models is
propelling an acceleration in late-time cosmic expansion.  However,
the physical nature of dark energy is not yet understood, and its
effects are subject to intense observational scrutiny.

Efforts in this area to date have focused on mapping out the cosmic
expansion history using baryon acoustic oscillations (BAOs) as a
standard ruler (e.g.\ Beutler et al.\ 2011, Blake et al.\ 2011b,
Anderson et al.\ 2014, Kazin et al.\ 2014, Delubac et al.\ 2015,
Aubourg et al.\ 2015, Alam et al.\ 2016) and Type Ia supernovae as
standard candles (e.g.\ Conley et al.\ 2011, Suzuki et al.\ 2012,
Betoule et al.\ 2014).  These probes have yielded important
constraints on the `homogeneous expanding Universe', including $\sim
1\%$ distance measurements and a $\sim 5\%$ determination of the value
of the equation-of-state of dark energy, $w$.  However, measurements
of the laws of gravity that describe the `clumpy Universe' are
currently less advanced, and only a combination of complementary
observations of expansion and gravitational growth will discriminate
between the different possible physical manifestations of dark energy.
Efforts have focused on establishing whether the laws of General
Relativity (GR), well-tested on solar-system scales, are a good
description of gravity on cosmological scales 14 orders of magnitude
larger.

There are two particularly important observable signatures of
gravitational physics that can be used for this purpose, and these two
methods gain considerable leverage when combined.  The first
observable is the peculiar motions of galaxies as they fall toward
overdense regions as non-relativistic test particles.  These motions
produce correlated Doppler shifts in galaxy redshifts that create an
overall clustering anisotropy as a function of the angle to the
line-of-sight, known as redshift-space distortion (RSD).  This pattern
has been measured by a number of galaxy redshift surveys (e.g.\ Blake
et al.\ 2011a, Beutler et al.\ 2012, de la Torre et al.\ 2013,
Samushia et al.\ 2014, Beutler et al.\ 2014, Marin et al.\ 2016, Alam
et al.\ 2016) and has permitted the growth rate of cosmic structure to
be measured with $\sim 10\%$ accuracy at some epochs.  The second
gravitational probe is the patterns of weak lensing imprinted by the
deflections of light rays from distant galaxies as they travel through
the intervening large-scale structure as relativistic test particles.
This signal may be measured using correlations in the apparent shapes
of background galaxies in deep imaging surveys (e.g.\ Heymans et
al.\ 2012, Huff et al.\ 2014, Kuijken et al.\ 2015, Becker et
al.\ 2016, Hildebrandt et al.\ 2016b).  Whilst the cosmological
parameter constraints possible from gravitational lensing statistics
are still improving, the measurement offers several key advantages
such as its insensitivity to galaxy bias.

Velocities and lensing are complementary because only their
combination allows general deviations to the Einstein field equations
to be constrained (Zhang et al.\ 2007, Song et al.\ 2011).  Modern
theories of gravity may be classified by the manner in which they warp
or perturb the spacetime metric (and the way this warping is generated
by matter).  In general two types of perturbations are possible:
spacelike and timelike.  In GR these perturbations are equal and
opposite, but in `modified gravity' scenarios a difference is
predicted.  Examples of such frameworks include generalizing the
`action' of GR as a function of the Ricci curvature, such as in $f(R)$
gravity models (Sotiriou \& Faraoni 2010), or embedding ordinary 3+1
dimensional space into a higher-dimensional manifold such as
`Cascading gravity' (de Rham et al.\ 2008) or `Galileon gravity' (Chow
\& Khoury 2009).  These scenarios make different observable
predictions.

Joint cosmological fits to weak gravitational lensing and galaxy
redshift-space distortion statistics can be performed using datasets
without sky overlap (e.g.\ Simpson et al.\ 2013).  However, the
availablity of overlapping imaging and spectroscopic surveys yields
several scientific benefits.  First, since the same density
fluctuations source both the lensing and galaxy velocity signals, the
partially-shared sample variance reduces the uncertainty in the
gravity fits (McDonald \& Seljak 2009), and the addition of the
shape-density correlation statistics (`galaxy-galaxy lensing') enables
new measurements to be constructed such as the `gravitational slip'
(Zhang et al.\ 2007).  A series of authors (Gaztanaga et al.\ 2012;
Cai \& Bernstein 2012; de Putter, Dore \& Das 2014; Kirk et al.\ 2015;
Eriksen \& Gaztanaga 2015) have predicted statistical improvements
resulting from overlapping surveys, although the degree of this
improvement depends on assumptions and survey configuration
(Font-Ribera et al.\ 2014).

Perhaps more importantly, the actual benefit of overlapping surveys
exceeds statistical forecasts because weak lensing measurements are
limited by a number of sources of systematic error which may be
mitigated using same-sky spectroscopic-redshift observations.  One of
the most significant systematic errors is the calibration of the
source photometric redshifts which are required for cosmic shear
tomography (Ma, Hu \& Huterer 2006).  Overlapping spectroscopic
surveys are a powerful means of performing this calibration (Newman et
al.\ 2015), using approaches including both observation of complete
spectroscopic sub-samples and analysis of cross-correlation statistics
(McQuinn \& White 2013, de Putter, Dore \& Das 2014).  Conversely, the
gravitational lensing imprint allows independent calibration of the
galaxy bias parameters that are a key systematic limitation to
redshift-space distortion analysis (e.g.\ Buddendiek et al.\ 2016).
Finally, overlapping imaging and spectroscopy enables a wide range of
other science including studies of galaxy clusters, strong lensing
systems and galaxy evolution.

The first wide-area overlapping spectroscopic and cosmic shear surveys
only recently became available\footnote{The Sloan Digital Sky Survey
  is a shallow lensing-spectroscopy survey that has previously allowed
  some measurements of this type (e.g.\ Reyes et al.\ 2010, Mandelbaum
  et al.\ 2013), but it suffers from significant levels of lensing
  systematics (Huff et al.\ 2014) such that cosmic shear studies are
  not permitted outside the Stripe 82 area.  Deep, narrow redshift
  surveys with lensing overlap also exist.} and currently span a
shared area of $\sim 500$ deg$^2$, consisting of an overlap between
two lensing imaging surveys -- the Canada-France-Hawaii Telescope
Legacy Survey (CFHTLS; Gwyn 2012, Heymans et al.\ 2012) and the 2nd
Red Sequence Cluster Survey (RCS2; Gilbank et al.\ 2011, Hildebrandt
et al.\ 2016a) -- and two spectroscopic redshift surveys -- the WiggleZ
Dark Energy Survey (Drinkwater et al.\ 2010) and the Baryon
Oscillation Spectroscopic Survey (BOSS; Dawson et al.\ 2013).  This
overlap has permitted a number of studies including a new consistency
test of GR via a measurement of gravitational slip at $z=0.6$ (Blake
et al.\ 2016), joint constraints on halo occupation distribution and
cosmological parameters (More et al.\ 2015), tests of imaging
photometric redshift performance via cross-correlation (Choi et
al.\ 2016) and new measurements of small-scale galaxy bias parameters
(Buddendiek et al.\ 2016).

Wide-area overlap between spectroscopic and imaging surveys requires
significant further extension to realize its full scientific
potential.  Two of the deep imaging surveys currently being performed
to measure gravitational lensing -- the 1500 deg$^2$ Kilo-Degree
Survey (KiDS; Kuijken et al.\ 2015) at the European Southern
Observatory VLT Survey Telescope (VST), and the 5000 deg$^2$ Dark
Energy Survey (DES; Becker et al.\ 2016) at the Blanco Telescope --
are located largely in the southern hemisphere, whereas the largest
existing wide-area spectroscopic surveys have been carried out by the
Sloan Telescope in the northern hemisphere.\footnote{A third
  in-progress deep-imaging lensing survey, using the Hyper-Suprime
  Camera (HSC) at the Subaru telescope, is mapping an area similar to
  KiDS with greater depth and will benefit from overlap with BOSS.}
With this in mind, we have created the 2-degree Field Lensing Survey
(2dFLenS)\footnote{Our website is {\tt http://2dflens.swin.edu.au}}, a
new southern-hemisphere spectroscopic redshift survey using the
Anglo-Australian Telescope (AAT).  The 2dF-AAOmega multi-fibre
spectroscopic system at the AAT has conducted a series of such
projects including the 2-degree Field Galaxy Redshift Survey (2dFGRS;
Colless et al.\ 2001), the WiggleZ Dark Energy Survey (Drinkwater et
al.\ 2010), the Galaxy And Mass Assembly survey (GAMA; Driver et
al.\ 2011), and OzDES (Yuan et al.\ 2015).

This paper describes the design, performance and initial clustering
analysis of 2dFLenS.  Key initial scientific analyses, some in
conjunction with KiDS, are presented by five associate papers (Joudaki
et al.\ 2016b, Johnson et al.\ 2016, Amon et al.\ 2016, Wolf et
al.\ 2016, Janssens et al.\ 2016).  Section \ref{secdesign} motivates
the survey design: the choice of fields and targets.  Section
\ref{sectarsel} describes the process of selecting targets from input
photometric imaging catalogues, and Section \ref{secspecobs} discusses
the spectroscopic observing campaign including AAT data reduction and
galaxy redshift determination.  Section \ref{secselfunc} describes the
calculation of the selection function of the spectroscopic
observations, which forms the basis of the ensuing galaxy clustering
measurements.  Section \ref{secmock} outlines the construction of the
survey mock catalogues which are used to estimate the covariance
matrix of the clustering statistics, whose measurement is discussed in
Section \ref{secclus}.  We summarize in Section \ref{secconc}.

\section{Survey design}
\label{secdesign}

\subsection{Choice of fields}

The purpose of 2dFLenS is to extend the coverage of
spectroscopic-redshift observations that overlap with deep optical
imaging surveys performed to measure weak gravitational lensing.  The
principal focus of our new spectroscopic coverage is the area being
imaged by the Kilo-Degree Survey\footnote{\tt
  http://kids.strw.leidenuniv.nl} (KiDS; de Jong et al.\ 2015, Kuijken
et al.\ 2015), a new lensing dataset in the southern sky.  The KiDS
footprint is planned to encompass 1500 deg$^2$, divided into two
approximately equal areas around the Southern Galactic Cap (SGC) and
Northern Galactic Cap (NGC).  Approximately 500 deg$^2$ of the KiDS
NGC survey region is already covered by deep spectroscopic data
provided by the Baryon Oscillation Spectroscopic Survey (BOSS; Dawson
et al.\ 2013).  The remaining KiDS area lacks deep, wide-area
spectroscopic coverage, although two shallower redshift surveys have
performed overlapping observations: the Galaxy And Mass Assembly
survey (GAMA; Driver et al.\ 2011) and the 2-degree Field Galaxy
Redshift Survey (2dFGRS; Colless et al.\ 2001).  However, neither of
these existing datasets has the depth nor coverage to address our
scientific aims.

In addition to the KiDS region, 2dFLenS also conducted observations in
sky areas covered by two other deep lensing imaging surveys: the
Canada France Hawaii Telescope Legacy Survey (CFHTLS; Gwyn 2012,
Heymans et al.\ 2012) and the 2nd Red Sequence Cluster Survey (RCS2;
Gilbank et al.\ 2011, Hildebrandt et al.\ 2016a).  In particular, we
targeted regions of those surveys which possessed either no or partial
deep spectroscopic follow-up: CFHTLS regions W1 and W2, and RCS2
regions 0320, 0357 and 1111.  The right ascension and declination
boundaries of all these fields are listed in Table \ref{tabregions}.

\begin{figure*}
\begin{center}
\resizebox{16cm}{!}{\rotatebox{270}{\includegraphics{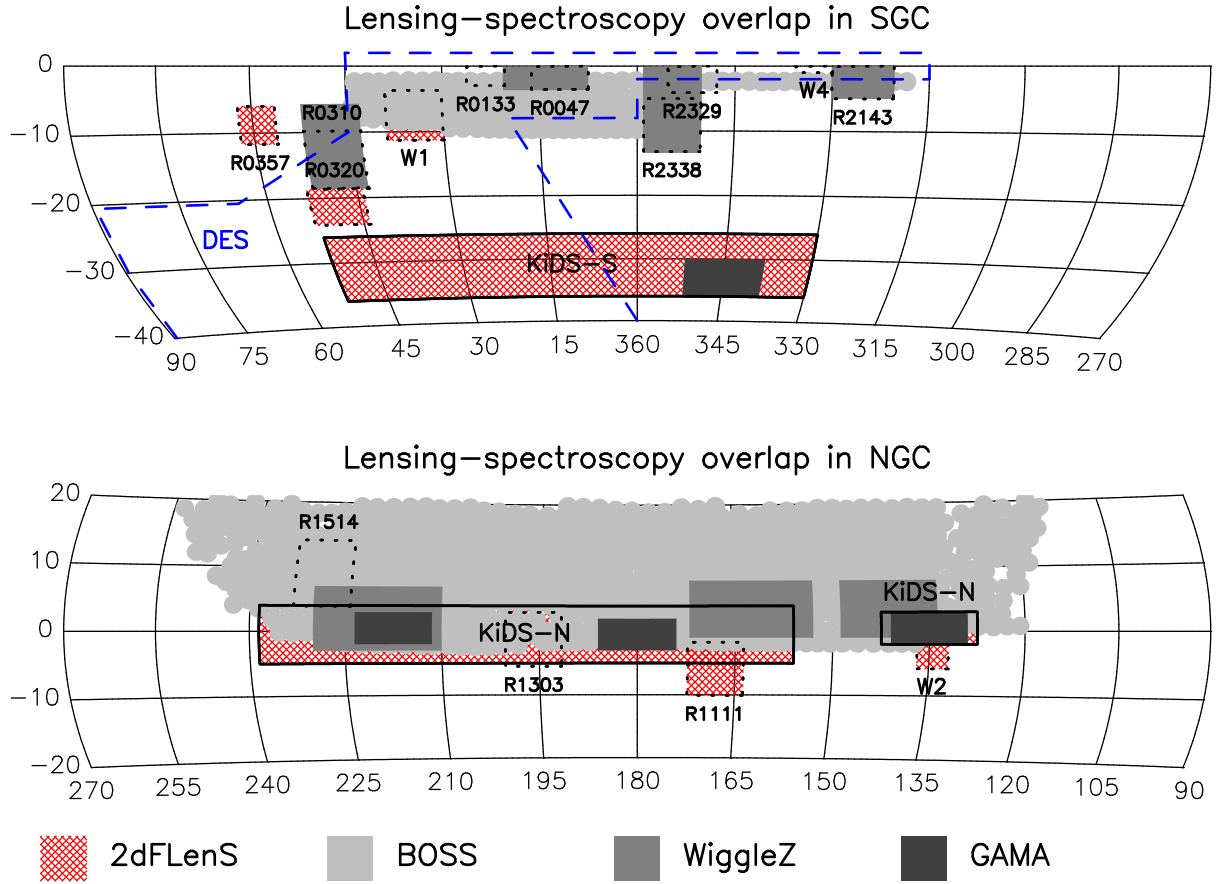}}}
\end{center}
\caption{Locations of current spectroscopic and imaging surveys in the
  Southern Galactic Cap (SGC) and Northern Galactic Cap (NGC).  Deep
  spectroscopic surveys (BOSS, WiggleZ and GAMA) are indicated as grey
  shaded regions.  Deep imaging surveys (CFHTLS, RCS2 and KiDS) are
  displayed as outlined rectangles, labelled by field names.  The
  approximate DES footprint is located inside the dashed blue line.
  The originally-planned 2dFLenS spectroscopic coverage is displayed
  using cross-hatched red shading and spans almost 1000 deg$^2$.}
\label{figoverlaps}
\end{figure*}

\begin{table}
\caption{Right Ascension and Declination boundaries in degrees of
  regions targeted for observation by 2dFLenS.  The top and bottom
  halves of the table list SGC and NGC regions, respectively.}
\begin{center}
\begin{tabular}{|c|c|c|c|c|}
\hline
Region & min R.A. & max R.A. & min Dec. & max Dec. \\
\hline
KiDS-S & 330.0 & 52.5 & -36.0 & -26.0 \\
RCS 0320 & 44.0 & 53.2 & -24.1 & -18.5 \\
RCS 0357 & 56.3 & 62.2 & -11.7 & -6.0 \\
CFHTLS W1 & 30.1 & 38.9 & -11.3 & -3.7 \\
\hline
KiDS-N (1) & 127.5 & 142.5 & -2.0 & 3.0 \\
KiDS-N (2) & 156.0 & 238.5 & -5.0 & 4.0 \\
RCS 1111 & 163.7 & 172.3 & -10.1 & -1.7 \\
CFHTLS W2 & 132.0 & 136.9 & -5.8 & -0.9 \\
\hline
\end{tabular}
\end{center}
\label{tabregions}
\end{table}

\begin{figure*}
\begin{center}
\resizebox{12cm}{!}{\rotatebox{270}{\includegraphics{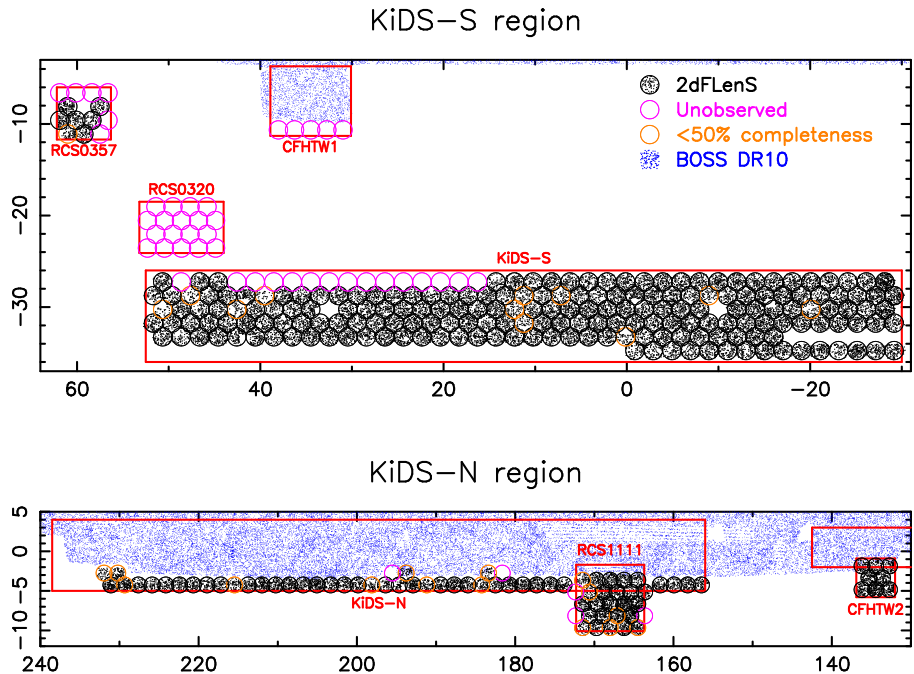}}}
\end{center}
\caption{The final coverage of 2dFLenS observations within the deep
  imaging survey regions outlined by the red rectangles.  The dark
  points display the locations of successful 2dFLenS galaxy redshifts,
  with BOSS galaxy redshifts indicated by the blue points.  The
  originally-planned 2dFLenS AAT pointings are displayed as the black
  circles, observations suffering low completeness at the end of the
  survey are indicated by orange circles, and unobserved pointings are
  highlighted in magenta.}
\label{figprogress}
\end{figure*}

Figure \ref{figoverlaps} illustrates the location of these regions in
more detail.  The cross-hatched red shaded area indicates the fields
originally planned to be targeted for observation by 2dFLenS,
extending existing coverage by BOSS.  This area comprised a total of
985 deg$^2$ (731 and 254 deg$^2$ in the SGC and NGC, respectively).
Our observations also overlap with the footprint of the Dark Energy
Survey (DES).  The final status of our spectroscopic campaign is
illustrated in Figure \ref{figprogress} and discussed in Section
\ref{secstatus}.

We tiled the 2dFLenS observation regions with 2-degree diameter
circular pointings of the 2dF+AAOmega spectroscopic system at the
Anglo-Australian Telescope (AAT), using a hexagonal pointing grid with
fixed field centres.  Usage of a fixed pointing grid, rather than an
adaptive, overlapping pointing grid, simplifies the determination of
the angular completeness of the observations via a ratio of successful
redshifts to intended targets computed in unique sectors.  In total we
defined 324 AAT pointing centres, 245 in the SGC and 79 in the NGC,
which were suitable for observation.  The distribution of these field
centres is displayed in Figure \ref{figprogress}.  We excluded a small
fraction of intended field centres which lacked appropriate input
imaging data as discussed in Section \ref{sectarsel}.

The 2dFLenS project applied for competitive time allocation at the AAT
in March 2014 and was allocated a total of 53 nights spread across the
14B, 15A and 15B semesters, the majority of which occurred in `grey
time' with regard to moon phase.

\subsection{Choice of targets}

The set of galaxies targeted for spectroscopic observation by
2dFLenS was selected to enable two principal scientific goals:
\begin{itemize}
\item Measurement of the gravitational lensing signal imprinted by the
  spectroscopic targets in the apparent shapes of background sources
  (`galaxy-galaxy lensing'), and the comparison of this lensing signal
  with the amplitude of galaxy peculiar velocities driven by the same
  density fluctuations, across a wide redshift range.
\item Determination of the source redshift distribution in the
  overlapping imaging survey, using both direct photometric-redshift
  calibration (enabled by spectroscopy of a complete sub-sample) and
  cross-correlation techniques (using the clustering between the
  imaging sources and the spectroscopic sample in narrow redshift
  slices).
\end{itemize}

The optimal choice of targets for the first goal, whose correlation
with background source shapes will maximize the resulting
galaxy-galaxy lensing signal, is Luminous Red Galaxies (LRGs), which
preferentially trace dense areas of the Universe and hence imprint the
strongest gravitational lensing signal.  Bright LRGs in the redshift
range $z < 1$ can be readily selected using well-understood colour and
magnitude cuts developed by previous observational projects such as
the Sloan Digital Sky Survey (SDSS, Eisenstein et al.\ 2001), the
2dF-SDSS LRG And Quasar survey (2SLAQ, Cannon et al.\ 2006), the
Baryon Oscillation Spectroscopic Survey (BOSS, Dawson et al.\ 2013)
and the Extended Baryon Oscillation Spectroscopic Survey (eBOSS,
Dawson et al.\ 2016).  We utilized these colour cuts, which are
inspired by the evolution with redshift of an early-type galaxy
template, in particular the 4000\AA\ spectral break, through the
optical filter system.  The majority of our survey area overlaps with
KiDS, which has a weighted mean redshift of $\sim 0.7$ (Kuijken et
al.\ 2015), where the weights reflect the accuracy of the weak lensing
shape measurement for each object.  We therefore prioritized
spectroscopic targets with $z < 0.7$ to maximise the number of lenses
that are in front of our source galaxies.

A disadvantage of targetting an LRG lens sample for our test of
gravitational physics is that its high galaxy bias factor, $b \approx
2$, results in a low redshift-space distortion (RSD) signal, whose
amplitude is determined by the parameter $\beta = f/b$ where $f$ is
the growth rate of cosmic structure.  The higher galaxy-galaxy lensing
signal, however, compensates for the lower RSD signal, rendering LRGs
the optimal choice of target for this scientific goal.

Turning now to the second goal: determination of the source redshift
distribution by cross-correlation mandates a spectroscopic sample
overlapping the imaging data across the widest possible redshift
range, but (unlike direct photometric-redshift calibration) is
agnostic regarding the spectroscopic sample's galaxy type, which is a
matter of observational convenience.  Given that our available
target-selection imaging is insufficiently deep for efficient
identification of high-redshift emission-line galaxies, we utilized
LRGs for this purpose as well.  The practical limitations of our
target-selection imaging, together with the integration time available
for our observations, restricted the accessible redshift range to $z <
0.9$.

Finally, photometric-redshift determination by direct calibration
requires the construction of complete spectroscopic-redshift datasets
spanning volumes sufficiently large to minimize the impact of sample
variance on this calibration (Cunha et al.\ 2012).  We therefore
selected a random sub-sample of galaxies to facilitate this set of
investigations, within a magnitude range defined by a faint limit ($r
\approx 19.5$) ensuring highly complete redshift determination in all
observing conditions, and a bright limit ($r \approx 17$) minimizing
overlap with current and future wide-area complete spectroscopic
samples such as 2dFGRS, SDSS and the Taipan Galaxy Survey\footnote{The
  Taipan Galaxy Survey {\tt (http://www.taipan-survey.org)} is a new,
  wide-area low-redshift spectroscopic survey scheduled to begin at
  the U.K.\ Schmidt Telescope in Australia in 2017.}.  Given that the
clustering of this magnitude-limited sample does not need to be
measured, it serves as an ideal set of `filler' targets which can be
prioritized below the LRGs scheduled for observation in each AAT
pointing, ensuring that all spectroscopic fibres are allocated.

In addition to our main target classes, we also included a set of
sparsely-distributed `spare fibre' targets within the 2dFLenS
observations.  These samples are described in Section
\ref{secspareselect}.

\section{Target selection}
\label{sectarsel}

\subsection{Imaging catalogues for target selection}

2dFLenS targets are selected using a variety of photometric
catalogues, depending on the sky area of observation.  The majority of
2dFLenS pointings are located within the KiDS survey footprint.
However, KiDS imaging observations were still ongoing when 2dFLenS
commenced, and therefore it was not possible to employ KiDS data for
2dFLenS target selection.  We instead used an overlapping shallower
and wider optical imaging survey, VST-ATLAS\footnote{\tt
  http://astro.dur.ac.uk/Cosmology/vstatlas} (Shanks et al.\ 2015),
for this purpose.  ATLAS is sufficiently deep for the selection of the
2dFLenS samples and enabled target selection across the majority of
the planned pointings within the KiDS footprint.  For convenience, we
also used ATLAS data to select targets in the RCS1111 region.  We
describe our processing of the ATLAS imaging data in the next
sub-section, followed by brief summaries of the CFHTLS and RCS2
imaging catalogues that we also employ for 2dFLenS target selection.

\subsubsection{VST-ATLAS imaging}
\label{secatlas}

The KiDS and ATLAS imaging surveys are both performed using the
OmegaCAM instrument at the European Southern Observatory (ESO) VLT
Survey Telescope (VST) and the same filter system.  OmegaCAM is an $8
\times 4$ CCD mosaic whose chips are $4102 \times 2048$-pixel arrays
which sample the focal plane at a uniform scale of $0.214$ arcsec.
VST-ATLAS is a `Sloan-like' imaging survey in the southern hemisphere
observed using five optical filters $ugriz$, with a limiting magnitude
$r \approx 22.5$, shallower than the KiDS limit of $r \approx 24$.
Relevant ATLAS survey properties are summarized in Table
\ref{tabatlas}.

\begin{table}
\caption{Characteristics of the co-added ATLAS imaging data used for
  2dFLenS target selection.  The columns indicate the exposure time in
  each of the 5 filters $ugriz$, together with the mean and standard
  deviation of the limiting AB magnitudes $m_{\rm lim}$ and seeing
  values across the ATLAS fields.  In this Table, the limiting
  magnitudes are defined as the 5-$\sigma$ detection limit within an
  annulus of radius 2 arcsec.}
\begin{center}
\begin{tabular}{llll}    
\hline
\multicolumn{1}{c}{Filter} & \multicolumn{1}{c}{expos. time [s]} &
\multicolumn{1}{c}{$m_{\rm lim}$ [AB mag]} & \multicolumn{1}{c}{seeing [$''$]}\\ 
\hline
$u$ & $2-4\times 60$ (120-240) & $21.97 \pm 0.21$ & $1.11 \pm 0.20$\\ 
$g$ & $2\times 50$ (100) & $23.04 \pm 0.12$ & $1.00 \pm 0.25$\\ 
$r$ & $2\times 45$ (90) & $22.46 \pm 0.20$ & $0.89 \pm 0.19$\\ 
$i$ & $2\times 45$ (90) & $21.79 \pm 0.23$ & $0.86 \pm 0.23$\\ 
$z$ & $2\times 45$ (90) & $20.65 \pm 0.23$ & $0.87 \pm 0.22$\\ 
\hline                  
\end{tabular}
\end{center}
\label{tabatlas}
\end{table}

The following is a short description of our ATLAS data processing for
2dFLenS target selection.  Our reduction starts with the raw OmegaCAM
data available at the ESO archive\footnote{{\tt
    http://archive.eso.org}.  ESO public source catalogues were not
  available at the start of our project.} at the time of processing
(initially 1 Dec 2013, updated 22 Dec 2014).  Our processing
algorithms are implemented in the publicly-available reduction
pipeline {\sc theli}\footnote{\tt http://www.astro.uni-bonn.de/theli}
and are described by Erben et al.\ (2005) and Schirmer (2013).  Our
{\sc theli} processing of ATLAS data consists of the following steps:

\begin{itemize}

\item We corrected for the significant cross-talk effects present in
  the three OmegaCAM CCDs in the left part of the uppermost row of the
  mosaic.

\item We removed the instrumental signature simultaneously for all
  data obtained in 2-week periods around each new-moon and full-moon
  phase, which define our processing runs (see Section 4 of Erben et
  al.\ 2005).  We assume here that the instrument configuration is
  stable within each processing run.  Division of data into these moon
  phases is convenient as it corresponds to the usage of certain
  filter combinations ($u$, $g$ and $r$ during new moon; $i$ and $z$
  during full moon).

\item First-order photometric zeropoints were estimated for each
  processing run using all images which overlap with SDSS Data Release
  10 (Ahn et al.\ 2014), assuming that photometric conditions were
  stable within the run.  We used between 30 and 150 such images with
  good airmass coverage for each processing run.

\item We subtracted the sky from all individual chips. These data form
  the basis for image co-addition in the final step.
 
\item We astrometrically calibrated the ATLAS imaging using the 2MASS
  catalogue (Skrutskie et al.\ 2006).

\item The astrometrically-calibrated data were co-added with a
  weighted-mean approach (see Erben et al.\ 2005).  The identification
  of pixels that should not contribute, and the pixel weighting of
  usable regions, is performed in the same manner as described by
  Erben et al.\ (2009, 2013) for CFHTLS data.

\item We did not apply an illumination correction to the imaging data,
  but implemented this correction to the catalogue magnitudes as
  described below.

\end{itemize}

We used a total of (680, 295) ATLAS pointings in the (SGC, NGC)
2dFLenS regions for target selection.  We generated a source catalogue
using the source extraction software {\sc SExtractor} (Bertin \&
Arnouts 1996) to analyze the co-added $r$-band image.  The selection
of the $r$-band as our detection band was motivated by the more
uniform quality of the ATLAS data in this band.  The alternative
$i$-band data is imaged in bright time and as such is more subject to
issues of scattered light during early VST imaging before baffling was
installed at the telescope.

Matched aperture photometry and colours were measured for the object
catalogue using {\sc SExtractor} in dual-extraction mode to analyze
PSF Gaussianised $u$, $g$, $r$, $i$, and $z$ images.  PSF
Gaussianisation across the 5-bands is achieved by modelling the
anisotropic PSF variation across each image followed by a convolution
with a spatially-varying kernel.  The resulting multi-band data has
identical Gaussian PSFs such that aperture magnitudes (defined by the
isophotes in our Gaussianised detection band) now measured flux from
the same region of the galaxy in each band.  The method we employ is
detailed in Hildebrandt et al.\ (2012).

Whilst our PSF Gaussianisation method provides an optimal measurement
of galaxy colour, it does not provide a total magnitude or `model
magnitude' in each band.  Accurate measurements of total magnitudes
can only be achieved through galaxy profile fitting.  A reasonable
approximation, however, is to use the {\tt MAG\_AUTO} measurement from
{\sc SExtractor} which employs a flexible elliptical aperture around
each object.  When measuring photometry in dual-extraction mode,
however, this measurement is only made in the detection band.  In
order to estimate total magnitudes in other bands we used the
difference between {\tt MAG\_AUTO} measured in the original detection
$r$-band, and the isophotal magnitude {\tt MAG\_ISO} measured by {\sc
  SExtractor} in the PSF Gaussianised $r$-band image, as a proxy for
the missed flux during the matched aperture photometry measurement,
such that
\begin{equation}
m_A = {\tt MAG\_AUTO\_r} + {\tt MAG\_ISO\_cor\_m} - {\tt MAG\_ISO\_r} ,
\end{equation}
where $m = \lbrace u, g, r, i, z \rbrace$ and {\tt MAG\_ISO\_cor}
includes a catalogue-based illumination correction.  This correction
was generated by a 2-dimensional polynomial fit to the zero-point
variation across the mosaic in each of the magnitude bands.  Dust
extinction corrections (Schlegel, Finkbeiner \& Davis 1998) were then
applied to the ATLAS magnitudes using
\begin{equation}
m_A \rightarrow m_A - {\tt EXTINCTION\_m} .
\end{equation}
All magnitudes were calibrated to the AB system.

\subsubsection{CFHTLS imaging}

CFHTLenS\footnote{\tt http://www.cfhtlens.org} (Heymans et al.\ 2012)
is a deep multi-colour imaging survey optimized for weak lensing
analyses, observed as part of the Canada-France-Hawaii Telescope
Legacy Survey (CFHTLS) in five optical bands $ugriz$, using the 1
deg$^2$ camera MegaCam.  The imaging data, which have limiting
5-$\sigma$ point-source magnitude $i \approx 25.5$, cover 154 deg$^2$
split into four fields, two of which (W1 and W4) already overlap with
deep spectroscopic data provided by BOSS.  2dFLenS observations
prioritized targetting of a third region, W2, as displayed in Figure
\ref{figoverlaps}.  Target selection in this area was performed using
the publicly-available CFHTLenS photometric catalogues (Erben et
al.\ 2013).

\subsubsection{RCS2 imaging}

The 2nd Red Sequence Cluster Survey (RCS2; Gilbank et al.\ 2011) is a
$\sim 800$ deg$^2$ imaging survey in three optical bands $grz$ also
carried out with the CFHT, with a limiting magnitude $r \approx 24.3$.
Around two-thirds of RCS2 has also been imaged in the $i$-band.  The
survey area is divided into 14 patches on the sky, each with an area
ranging from 20 to 100 deg$^2$.  Nine of these regions already overlap
with deep spectroscopic data provided by the BOSS and WiggleZ surveys
(Blake et al.\ 2016); 2dFLenS observations planned to target three
further areas as indicated in Figure \ref{figoverlaps}: RCS 0320, 0357
and 1111, although observations were only achieved in the last two of
these regions owing to poor weather and the prioritization of fields
overlapping KiDS.  For convenience we performed target selection using
ATLAS data in the RCS 1111 region.  Target selection in the other
areas was performed using the RCSLenS\footnote{\tt
  http://www.rcslens.org} photometric catalogues (Hildebrandt et
al.\ 2016a), a lensing re-analysis of the RCS2 imaging data performed
by applying the same processing pipeline as developed for CFHTLenS.

\subsubsection{WISE imaging}

The availability of infra-red data permits efficient star-galaxy
separation for high-redshift LRG selection (Prakash et al.\ 2015).  We
therefore matched our optical imaging catalogues with the AllWise
catalogue\footnote{\tt
  http://irsa.ipac.caltech.edu/cgi-bin/Gator/nph-scan?
  mission=irsa\&submit=Select\&projshort=WISE}.  We required sources
to have a good detection in $W1$ (${\tt w1snr} > 5$), and applied no
other selection flags.  We transformed the WISE magnitude ({\tt
  w1mpro}) to an AB magnitude using $W1 = {\tt w1mpro} + 2.683$, and
applied a Galactic dust extinction correction using $W1 \rightarrow W1
- 0.231 \, E_{B-V}$.

We matched objects between the optical and infra-red catalogues using
a search radius of 1.5 arcsec around each WISE source.  This search
area is small in comparison to the WISE beam of 6 arcsec, but was
found to be optimal through visual inspection of a sample of matched
galaxies.  Objects were matched when the WISE search radius was
contained within the ellipse defined by {\sc SExtractor} around the
corresponding optical source.  In the event that a WISE source was
matched to more than one optical source, the WISE photometry was
assigned to both optical sources and flagged as a blend.

\subsubsection{Magnitude transformations}

Since we aimed to reproduce various SDSS galaxy selections described
in Section \ref{seclrgselect}, and the filter systems used in our
input optical imaging surveys were not identical to SDSS filters, we
derived magnitude transformations between these surveys and SDSS using
an elliptical galaxy template spectrum.  These transformations are
detailed in Appendix \ref{secmagtrans}.

\subsubsection{Star-galaxy separation}
\label{secstargalaxy}

One of the main challenges in LRG selection is to separate stars from
galaxies, as the colour selection produces more stars than galaxies.
The quality of the ATLAS data, in terms of seeing, is significantly
better than SDSS, with an average seeing of $0.89$ arcsec in the
$r$-band (Table \ref{tabatlas}).  This high-quality data allows us to
separate stars from galaxies based on their size and shape.  We
performed a preliminary selection of stars based on a high-pass
detection threshold {\sc SExtractor} analysis of each $r$-band
exposure that enters the co-added image.  Candidate stars in each
exposure were identified on the stellar locus in the size-magnitude
plane.  Their ellipticity was then measured using the {\sc ksb}
algorithm (Kaiser, Squires \& Broadhurst 1995) and a
position-dependent model of the point-spread function (PSF) in each
exposure was derived iteratively, rejecting outlying objects with
non-PSF-like shapes from the sample.  Objects are defined to be stars
if they are identified as such in multiple exposures that enter the
co-added image.  This procedure provides a clean catalogue of
unresolved stellar objects which is removed from our LRG sample.
However, as our aim was to produce a clean galaxy sample, we also
imposed a further selection that the half-light radius of the object
{\tt FLUX\_RADIUS}, measured by {\sc SExtractor}, was greater than 0.9
times the measured seeing in the co-added image.

\subsubsection{Masks}

Image defects such as cosmic rays, saturated pixels, satellite tracks,
reflections and hot/cold pixels were recorded in a weight map image,
as described by Erben et al.\ (2013).  This map was incorporated in
the {\sc SExtractor} object detection analysis such that these defects
did not enter our source catalogue.  Additional stellar masks were
applied to remove diffraction spikes and `ghost' images around bright
stars.  These stellar masks were determined with the automated masking
algorithm described by Erben et al.\ (2009) and Kuijken et
al.\ (2015).  This uses standard stellar catalogues and knowledge
about the magnitude and positional dependence of the `ghosting' angle
for OmegaCAM.  Further masking of defects and image artefacts such as
spurious object detections, asteroids and satellite trails missed in
the weight map was performed through visual inspection as described in
Section \ref{seceyeballing}.

\subsubsection{Faint and bright magnitude limits}

In order to ensure that the optical magnitudes of ATLAS sources were
reliable for use in target selection, we imposed faint magnitude
limits $(25.2, 24.7, 24.1)$ in the $(g, r, i)$ bands.  These limits
where chosen as 1 standard deviation brighter in each band than the
mean total magnitude limit of the galaxy number counts in ATLAS fields
within the 2dFLenS survey region.  The equivalent limits we used for
RCSLenS and CFHTLenS data were $(27.1, 26.9, 26.4)$ and $(28.3, 27.7,
27.6)$.  Furthermore, due to image saturation we applied bright
magnitude limits $(17.4, 18.1, 17.8, 16.8)$ for RCSLenS in the $(g, r,
i, z)$ bands.  These limits corresponded to the faintest magnitude
that becomes saturated in any of the RCS images that overlap with the
2dFLenS survey region.  We therefore lost some brighter targets from
our selection, but as the RCS data is a single long-exposure image,
there is no alternative for obtaining reliable fluxes for these bright
targets in the RCS fields.

\subsubsection{Visual target inspection}
\label{seceyeballing}

We developed a web-based interface for visually inspecting
multi-wavelength postage stamp images of all LRGs selected for
observation by 2dFLenS.  Targets were removed from the sample if there
was clear evidence that they were artefacts or stars, or that their
apparent colours were influenced by nearby stars.  Our
multi-wavelength `cut-outs' server code repository is publicly
available\footnote{\tt https://github.com/dklaes/cutout\_server}.

\subsection{LRG sample selection}
\label{seclrgselect}

The LRG target selection for 2dFLenS employed similar colour and
magnitude cuts to those utilized by the SDSS, BOSS and eBOSS surveys,
in terms of transformed magnitudes in Sloan filters $\lbrace u_S, g_S,
r_S, i_S, z_S \rbrace$ (see Appendix \ref{secmagtrans} for the details
of the transformations).  We followed the evolution with redshift of
the LRG spectrum by defining separate colour cuts for selecting
low-redshift, mid-redshift and high-redshift 2dFLenS samples, matching
the surface density of AAOmega fibres over redshift range $z < 0.9$.

These selections make use of the colour variables
\begin{eqnarray}
& & c_\parallel = 0.7 \, (g_S-r_S) + 1.2 \, (r_S-i_S-0.18) , \label{eqcpar} \\
& & c_\perp = (r_S-i_S) - (g_S-r_S)/4 - 0.18 , \label{eqcperp} \\
& & d_\perp = (r_S-i_S) - (g_S-r_S)/8 . \label{eqdperp}
\end{eqnarray}
These variables define a convenient co-ordinate system for the locus
of early-type galaxies in the $g_S - r_S$ vs.\ $r_S - i_S$ plane, with
$c_\parallel$ increasing parallel to this track, and $c_\perp$
defining the distance perpendicular to the locus (Eisenstein et
al.\ 2001).  Cuts above lines of constant $d_\perp$ select early-type
galaxies at increasingly high redshift (Cannon et al.\ 2006).

\subsubsection{Low-redshift sample}

\begin{figure}
\begin{center}
\resizebox{8cm}{!}{\rotatebox{270}{\includegraphics{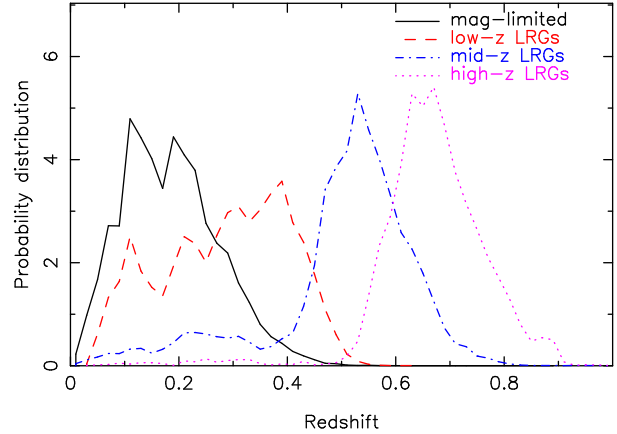}}}
\end{center}
\caption{Distribution of successful redshifts within each 2dFLenS
  target class.  The black solid, red dashed, blue dot-dashed and
  magenta dotted lines show the magnitude-limited, low-$z$, mid-$z$
  and high-$z$ LRG samples, respectively.}
\label{fignztype}
\end{figure}

First, we included galaxies satisfying `Cut I' or `Cut II' in the SDSS
LRG sample (Eisenstein et al.\ 2001), where `Cut I' is defined by
\begin{eqnarray}
& & 16.0 < r_S < 19.2 , \\
& & r_S < 13.1 + c_\parallel/0.3 , \\
& & |c_\perp| < 0.2 ,
\end{eqnarray}
and `Cut II' is defined by
\begin{eqnarray}
& & 16.0 < r_S < 19.5 , \\
& & c_\perp > 0.45 - (g_S - r_S)/6 , \\
& & g_S - r_S > 1.3 + 0.25 \, (r_S - i_S) .
\end{eqnarray}
We supplemented this sample with additional objects fulfilling the
BOSS `LOWZ' selection (Dawson et al.\ 2013):
\begin{eqnarray}
& & 16.0 < r_S < 19.6 , \\
& & r_S < 13.5 + c_\parallel/0.3 , \\
& & |c_\perp| < 0.2 .
\end{eqnarray}
These cuts are designed to isolate the locus of early-type galaxies in
colour space.  Low-$z$ LRG targets must be classified as galaxies by
the star-galaxy separation algorithm described in Section
\ref{secstargalaxy} (the fraction of stars targeted is $\sim 1\%$).
Figure \ref{fignztype} displays the redshift distribution of targets
selected in each 2dFLenS sample.  The low-$z$ LRGs span redshift range
$0.05 < z < 0.5$; the mean and standard deviation of the redshift
distribution is $0.29 \pm 0.12$.

\subsubsection{Mid-redshift sample}

The mid-$z$ LRG sample in 2dFLenS was selected using magnitude and
colour cuts similar to those employed by the BOSS `CMASS' sample
(Dawson et al.\ 2013):
\begin{eqnarray}
& & 17.5 < i_S < 19.9 , \\
& & r_S - i_S < 2 , \\
& & d_\perp > 0.55 , \\
& & i_S < 19.86 + 1.6 \, (d_\perp-0.8) .
\end{eqnarray}
A mid-$z$ LRG target must be classified as a galaxy by the star-galaxy
separation algorithm, and not already be selected for the low-$z$ LRG
sample.  The redshift distribution of mid-$z$ LRGs is displayed as the
blue dot-dashed line in Figure \ref{fignztype}.  The majority of the
objects are distributed in the redshift range $0.4 < z < 0.8$, with a
tail to lower redshifts $z < 0.4$.  The mean and standard deviation of
the redshift distribution is $0.50 \pm 0.14$.  These values are
comparable to those obtained by BOSS-CMASS, although the target
densities of the two surveys are somewhat different, as discussed
below.

\subsubsection{High-redshift sample}

The high-$z$ LRG sample in 2dFLenS was selected using joint optical and
infra-red magnitude and colour cuts (Prakash et al.\ 2015) similar to
those used to define the eBOSS LRG sample (Dawson et al.\ 2016):
\begin{eqnarray}
& & (r-W1) > 2 \, (r-i) , \label{eqselhiz} \\
& & r-i > 0.98 , \\
& & i-z > 0.6 , \\
& & 19.9 < i < 21.8 , \\
& & z < 19.95 .
\end{eqnarray}
A high-$z$ LRG target must not already be in the 2dFLenS low-$z$ or
mid-$z$ samples.  We do not apply size-based star-galaxy separation to
this sample; the optical-infrared colour cut in Equation
\ref{eqselhiz} is very effective for this purpose (Prakash et
al.\ 2015).  The redshift distribution of high-$z$ LRGs is displayed
as the magenta dotted line in Figure \ref{fignztype}.  The high-$z$
LRGs span redshift range $0.5 < z < 0.9$; the mean and standard
deviation of the redshift distribution is $0.67 \pm 0.10$.

\subsubsection{Size cut}
\label{secsize}

The resulting catalogue of selected LRG galaxies was larger than the
number of available 2dF-AAOmega fibres.  Given this, and in order to
homogenize the target density in the presence of variable seeing, we
applied a size cut to the low-$z$ and mid-$z$ LRG samples such that
${\tt FLUX\_RADIUS} > 4 \, {\rm pixels} = 0.86 \, {\rm arcsec}$.  We
do not apply this cut to the high-$z$ LRG sample.

\subsubsection{Target densities}

The average target densities of the (low-$z$, mid-$z$, high-$z$) LRG
samples selected across the 975 ATLAS fields were $(29, 65, 32)$
deg$^{-2}$.\footnote{These figures include sources which were later
  removed following visual inspection as described in Section
  \ref{seceyeballing}.}  The total density was therefore a good match
to the density of AAT fibres on the sky ($\approx 120$ deg$^{-2}$).
The target densities in the RCS2 and CFHTLS regions were similar for
the low-$z$ and mid-$z$ LRG samples, but approximately twice as high
for the high-$z$ sample due to the deeper optical data.  For
comparison, the target densities in the SDSS (BOSS-LOWZ, BOSS-CMASS,
eBOSS-LRG) samples are $(54, 94, 60)$ deg$^{-2}$ (Anderson et
al.\ 2014, Dawson et al.\ 2016) and therefore our LRG samples are
roughly a factor of 2 less dense than SDSS.  This difference is driven
by a combination of the size cut (Section \ref{secsize}), and
uncertainties in photometric calibration (Section \ref{secphotsys}).

\subsection{Magnitude-limited sample selection}

\subsubsection{Extended-source sample}

The extended-source magnitude-limited sample, designed to facilitate
direct source classification and photo-$z$ calibration, was selected
by randomly sub-sampling objects in the optical catalogues subject to
the following rules:
\begin{itemize}
\item Targets are restricted to the magnitude range $17 < r < 19.5$.
\item In order to increase the number of bright galaxies in the sample
  given the steepness of the source counts, the probability of
  selecting a target was increased by a factor of 2 with every
  magnitude brighter.
\item If the randomly-chosen object was already selected in another
  target class, this information was stored and the target was also
  included in the complete sample.
\item Objects were classified as galaxies by the star-galaxy
  separation algorithm described in Section \ref{secstargalaxy}.
\end{itemize}
Magnitude-limited targets were assigned lower priority than LRG targets
when allocating fibres in each field, such that the number varied in
anti-correlation with the angular clustering of the LRG sample.

\subsubsection{Point-source sample}

In the 15B semester, a new set of photo-$z$ calibration targets was
included in 2dFLenS observations.  By checking the star-galaxy
separation, we realized that the point-source sample clearly contained
unresolved galaxies, which we did not want to miss for the direct
photo-$z$ calibration.  We further wanted to measure the object class
composition of objects with colours which did not clearly correspond
to isolated single stars, such as QSOs, hot subdwarfs and white
dwarfs, M-dwarf/white-dwarf binaries and objects with apparently
unusual colours.  We thus added to the target catalogue
randomly-selected objects from the point-source sample, whose colours
did not clearly indicate a regular FGKM star (see Wolf et al.\ 2016).

\subsection{Spare fibre sample selection}
\label{secspareselect}

Wide-area spectroscopic surveys allow efficient follow-up of rare,
sparsely-distributed classes of objects whose spectra would be
difficult to obtain otherwise.  We included several such samples in
the 2dFLenS target pool.

\subsubsection{Red nugget sample}

The red nugget spare-fibre sample comprised Early-Type Galaxies (ETGs)
at $z < 1$ with effective radii $R_e$ and stellar masses $M_*$ similar
to compact ETGs at $z \sim 2$ ($\log{M_*} > 11$ and $R_e < 2$ kpc; van
Dokkum et al.\ 2008, Damjanov et al.\ 2009).  We therefore chose
targets that satisfied the low-$z$ or mid-$z$ LRG colour cuts but
failed the size-based star-galaxy separation described in Section
\ref{secsize}.  Instead, we used the optical-infrared colour for
star-galaxy separation (Equation \ref{eqselhiz}), supplementing the
main 2dFLenS LRG sample with objects that possessed LRG colours but
would be classified as stars based on size.  Since the high-$z$ LRG
sample already used optical-infrared colour, red nugget targets were
not added to this sample.  All targets were eyeballed to remove
objects affected by artefacts or close neighbours which may
contaminate the WISE photometry.  A total of $631$ unique red nugget
spectra were observed.

The abundance of red nuggets at $z < 1$ remains controversial.  Using
SDSS, Taylor et al.\ (2010) found zero ETGs in the redshift range
$0.066 < z < 0.12$ that had sizes and masses comparable to red nuggets
at $z \sim 2$.  However, in the WINGS ($0.04 < z < 0.07$; Valentinuzzi
et al.\ 2010) and PM2GC ($0.03 < z < 0.11$; Poggianti et al.\ 2013)
surveys of low-redshift clusters, a couple of hundred red nugget
analogues were found with number densities comparable to that of red
nuggets at $z \sim 2$.  In the COSMOS field, the number density of
compact ETGs is also similar to that observed at high redshift and
remains constant in the range $0.2 < z < 0.8$ (Damjanov et al.\ 2015).
The 2dFLenS red nugget sample will provide another measurement of the
number density and a significant increase in the number of $z < 1$ red
nuggets with spectra.

\subsubsection{Other samples}

We observed a small number of other spare-fibre targets, drawn from
pools of strong gravitational lensing system candidates and Brightest
Cluster Galaxies selected from the XMM Cluster Survey (XCS, Mehrtens
et al.\ 2012) and the South Pole Telescope (SPT, Bleem et al.\ 2015)
datasets.

\subsection{Flux calibrator sample selection}

Where $u$-band optical imaging data was available, three flux
calibrators per AAT field were included in the 2dFLenS sample using an
F-star selection (Yanny et al.\ 2009):
\begin{eqnarray}
& & -0.7 < 0.91 \, (u_S-g_S) + 0.415 \, (g_S-r_S) - 1.28 \nonumber \\
& & < -0.25 , \\
& & 0.4 < u_S-g_S < 1.4 , \\
& & 0.2 < g_S-r_S < 0.7 , \\
& & 17 < g_S < 18 .
\end{eqnarray}
Flux calibrators must also be classified as stars by the star-galaxy
separation algorithm described in Section \ref{secstargalaxy}.  We
used the F-star spectra (where available) during the data reduction
process to determine a mean sensitivity curve and zero-points for flux
calibration of 2dFLenS spectra.

\subsection{Photometric calibration challenges}
\label{secphotsys}

When the VST surveys were conceived, ATLAS and KiDS were designed to
facilitate precision-level photometry through overlap matching; the
tiling strategies include a half-field-of-view shift between the two
surveys.  KiDS and ATLAS were anticipated to be observed in parallel
with matched data acquisition rates such that the surveys could be
used in tandem for high-precision photometry.  However, in practice
ATLAS progress has greatly exceeded that of KiDS.  Therefore, given
that the overlapping area between ATLAS pointings is insufficient to
calibrate the photometric variation between fields, we are left with a
significant challenge to improve ATLAS photometric calibration beyond
the first-order process described in Section \ref{secatlas}.

Shanks et al.\ (2015) advocate using APASS data to improve
zeropoint-calibration for ATLAS beyond the ESO nightly standards.  We
investigated this approach, but found that the low number of objects
in the APASS catalogues that were unsaturated in the ATLAS imaging was
unlikely to improve our zeropoint-calibration beyond our already
improved SDSS-calibration.  Shanks et al.\ (2015) also advocate using
stellar locus regression (SLR) in colour-colour space, which we also
investigated.  Here we compared the colours of Pickles (1998) standard
stars to the colours of objects selected in our clean stellar
catalogue (described in Section \ref{secstargalaxy}).  We derived
linear shifts in colour to minimise the offset between the two
distributions.  Whilst this provided accurate colours for all objects
in the catalogue, using SLR to improve the accuracy of measured
magnitudes requires fixing of the photometry in one `pivot' band.  We
chose this pivot band in each field by minimizing the variance in the
linear offsets applied to the other bands.  Whilst on average this is
the most optimal method to select a pivot band, on an individual field
basis, this could well be the wrong choice.

The choice of an incorrect `pivot' band can have a significant impact
on LRG target selection, which depends on both colour and magnitude.
We therefore decided only to apply SLR corrections to fields that were
significant outliers in terms of the average value of $c_\perp$
(Equation \ref{eqcperp}) measured for galaxies with $16<r<19.6$.  We
selected $2\%$ of the fields that have an average $c_\perp$ more than
$2\sigma$ away from the mean $c_\perp$ for the full survey.  The
application of our SLR magnitude correction resulted in an acceptable
average value of $c_\perp$ for all but two of these fields, both of
which were found to have a high level of artefacts which required
manual masking.

Despite our efforts to achieve a good photometric calibration for
ATLAS data, a full solution will require a joint re-analysis with
KiDS, which has only recently acquired sufficient areal coverage to
facilitate this process.  For now, our ATLAS dataset still contains
significant photometric systematics which affect our LRG target
selection.  In Section \ref{secclussys} we describe the mitigation of
these effects in our clustering analysis, in which we are able to
marginalize over these systematics with minimal impact on our
scientific results.\footnote{In our direct photo-$z$ calibration study
  (Wolf et al.\ 2016) we partially correct for these effects by using
  WISE W1 photometry as a pivot band.  This approach produced improved
  results suitable for that study, but was not able to remove the
  clustering systematics completely.}

\section{Spectroscopic observations}
\label{secspecobs}

\subsection{2dF-AAOmega system and observing set-up}

The 2dFLenS observational project was performed at the
Anglo-Australian Telescope using the 2dF-AAOmega instrumentation.  The
2dF system (Lewis et al.\ 2002) is a multi-fibre positioner consisting
of two field plates mounted at the AAT prime focus, whose position may
be exchanged using a tumbling mechanism.  Whilst observations are
performed using one plate, fibres for the subsequent observation may
be configured on the other plate using a robot positioner.  A maximum
of 392 science fibres and 8 guide fibre bundles can be positioned over
a circular field-of-view with a diameter of two degrees.  The angular
diameter of each fibre on the sky is 2 arcsec.  The physical size of
the magnetic buttons on which fibres are mounted implies that fibres
cannot be positioned more closely together than 30 arcsec, and the
probability of successfully allocating fibres to each member of a pair
of galaxies decreases with pair separations below 2 arcmin.

Optical fibres (of length 38m) carry the light from the telescope
prime focus to the AAOmega spectrograph.  AAOmega is a bench-mounted
spectrograph consisting of blue and red arms split by a dichroic
(Saunders et al.\ 2004, Sharp et al.\ 2006).  2dFLenS utilizes the
580V and 385R AAOmega volume phase holographic gratings in the blue
and red arms respectively, providing a uniform resolving power of $R
\approx 1300$.  The total wavelength range of each observation was
3700 to 8800\AA, and we used the standard AAOmega dichroic with a
wavelength division of 5700\AA.

For each observation, target `field files' were prepared consisting of
the positions of science targets (with assigned priorities), potential
fiducial (guide) stars to align the field accurately, and potential
blank sky positions to sample the sky background to be subtracted
during data reduction.  The 2dF {\sc configure} software was used to
generate configuration files from these target lists.  This software
allocates the fibres using a simulated annealing algorithm (Miszalski
et al.\ 2006), such that all targets in each successive priority band
are preferentially allocated, and outputs a configuration file which
was utilized by the 2dF positioner.

The 2dFLenS field files for each pointing typically consisted of 600
science targets, 100 potential guide stars and 100 blank sky
positions, of which a subset of approximately 360, 8 and 25,
respectively, are allocated for observation.  In the software
configuration process we used the following science target priorities
(highest to lowest): flux calibrators (${\tt priority}=9$), spare
fibre targets (8), low-$z$ LRGs (7), mid-$z$ LRGs (6), high-$z$ LRGs
(5), and magnitude-limited sample (4).  In practice, the numbers in
each priority band imply that all targets with ${\tt priority} \ge 6$
and most targets with ${\tt priority} = 5$ were observed.  Flux
calibrators and spare fibre targets constitute small samples
(typically 3 and 5 objects per field, respectively), and were
therefore accorded the highest priority to ensure they were observed.

Fibre placement by the 2dF robot positioner requires $\sim 40$ minutes
for each field, for typical 2dFLenS configuration geometries.  This
duration specifies the minimum integration time for the observations.
In order to maximize the areal coverage of the survey we fixed the
integration time of each observation close to this limit: 45 minutes
split into three 15-minute exposures to assist with cosmic-ray
rejection.  The observing sequence for each telescope pointing also
included a standard fibre flat field (with exposure time 7 seconds)
and arc exposure for wavelength calibration (45 seconds).  Including
field acquistion, CCD read-out and other overheads, observations of
each 2dFLenS pointing could be completed in 1 hour.  We also acquired
`dome flat' fields for calibration purposes, whose use is described in
Section \ref{secdatared}.

\subsection{Guide star and blank sky positions}

We selected guide stars for 2dFLenS observations from 2MASS (Skrutskie
et al.\ 2006), to which the astrometry of our input imaging catalogues
is tied.  A cross-match with the UCAC4 catalogue (Zacharias et
al.\ 2013) was used to check that potential guide stars have
acceptably low proper motion and magnitudes within an appropriate
range.  In detail, guide stars satisfied the following criteria:
\begin{itemize}
\item UCAC4 $f$-band magnitude in the range $13.55 < m_f < 14.5$
\item Error in this magnitude less than $0.3$ mag
\item Measured proper motion $< 0.02$ arcsec yr$^{-1}$
\item Positional uncertainty $< 0.1$ arcsec
\item Offset in 2MASS-UCAC4 match $< 0.5$ arcsec
\end{itemize}
All guide star candidates were visually inspected using the web-based
interface described in Section \ref{seceyeballing}, and only utilized
if there was clear evidence that they were not galaxies, did not have
close companions and were located at the expected co-ordinates.

We determined potential blank sky locations by sampling random
positions from our optical images where the position satisfied the
joint criteria of containing no flux (as defined by an {\sc
  SExtractor} segmentation image using conservative parameters ${\tt
  MINAREA}=2$, ${\tt THRESH}=2$ and ${\tt ANALYSIS\_THRESH}=2$) and
being located at least 50 pixels (11 arcsec) from a stellar halo mask
or similar flag.  The minimum distance between potential sky postions
was specified as 5 pixels.

\subsection{Data reduction}
\label{secdatared}

\begin{figure*}
\begin{center}
\resizebox{12cm}{!}{\rotatebox{270}{\includegraphics{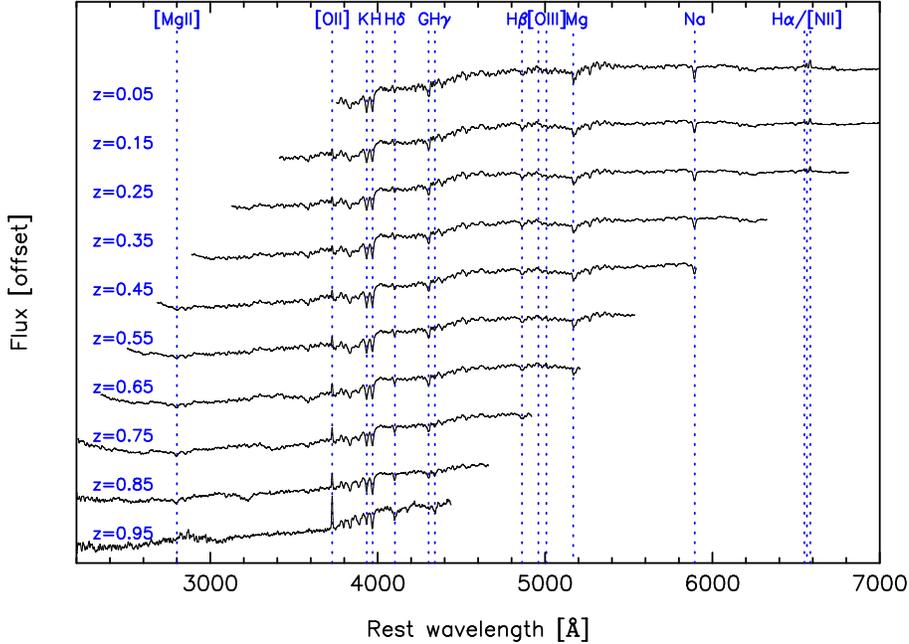}}}
\end{center}
\caption{Spectra of 2dFLenS LRGs with good-quality redshifts stacked
  by rest-frame wavelength in $\Delta z = 0.1$ slices.  Prominent
  spectral features are indicated by the vertical dotted lines.}
\label{figstackspec}
\end{figure*}

The AAOmega data were reduced during each observing run using the {\sc
  2dfdr} software developed at the AAO to process the science,
flat-field and arc frames.  The data from the blue and red
spectrograph arms for each field were reduced separately, and then
spliced together into a final complete spectrum.  We refer the reader
to Lidman et al.\ (2016) for a full description of the standard data
reduction process, and restrict our discussion here to one important
modification: in addition to the flat-field frames that are taken with
the flaps that fold in front of the 2dF corrector, we also acquired
flats using a patch on the windscreen that is painted white.  We refer
to the former as `flap flats' and the latter as `dome flats'.

As is standard practice in processing data taken with AAOmega, we used
the flap flats to measure the trace of the fibres on the CCDs (the
so-called `tramline map') and to determine the profile of the
fibres. We did not use the dome flats directly, since the
signal-to-noise ratio in the blue is too low, but instead used them to
correct the flap flats.

In more detail, we processed the dome flats and the flap flats in an
identical manner and then divided the flap flat by the dome flat. The
result was smoothed and then multiplied back into the flap flat. This
procedure preserved the high signal-to-noise ratio of the flap flat
while correcting the wavelength-dependent response of the flap flat.
The technique of using one kind of flat to correct another is commonly
used to process imaging data, and is often referred to as an
illumination correction.

The dome flats were taken once per run for each 2dF plate. We found
that acquiring dome flats more often, or for every set-up, did not
result in significantly better results, since the dome flats are very
stable once the absolute normalisation of fibre throughput is removed.

Using illumination corrections leads to improved data reduction
quality.  Systematic errors in the sky subtraction are significantly
smaller, especially when the background is high, which can occur
during nights when the moon is above the horizon (most of the 2dFLenS
data were taken during grey time). This then allows one to splice the
red and blue halves of the spectrum more accurately.  Errors in the
splicing can lead to a discontinuity in spectra (the so-called
dichroic jump) at this wavelength.  While there were several factors
that led to this discontinuity, the poor illumination offered by the
flap flats was the largest contributing factor.

Whilst further improvements in data reduction are possible
(e.g.\ better modelling of the fibre profile and scattered light), the
quality of the reduced data is sufficient for analyses requiring an
accurate estimate of the continuum such as equivalent widths, in
addition to measuring line fluxes and line centroids.

\subsection{Redshift determination}

\begin{figure}
\begin{center}
\resizebox{8cm}{!}{\rotatebox{270}{\includegraphics{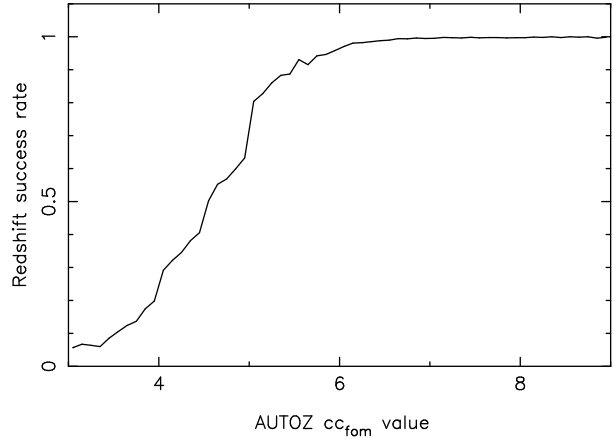}}}
\end{center}
\caption{Redshift success rate of 2dFLenS observations, defined by the
  fraction of spectra with redshift quality flag $\ge 3$, as a
  function of the cross-correlation parameter ${\rm cc}_{\rm fom}$
  determined by the {\sc autoz} code.}
\label{figredcompcc}
\end{figure}

\begin{figure*}
\begin{center}
\resizebox{16cm}{!}{\rotatebox{270}{\includegraphics{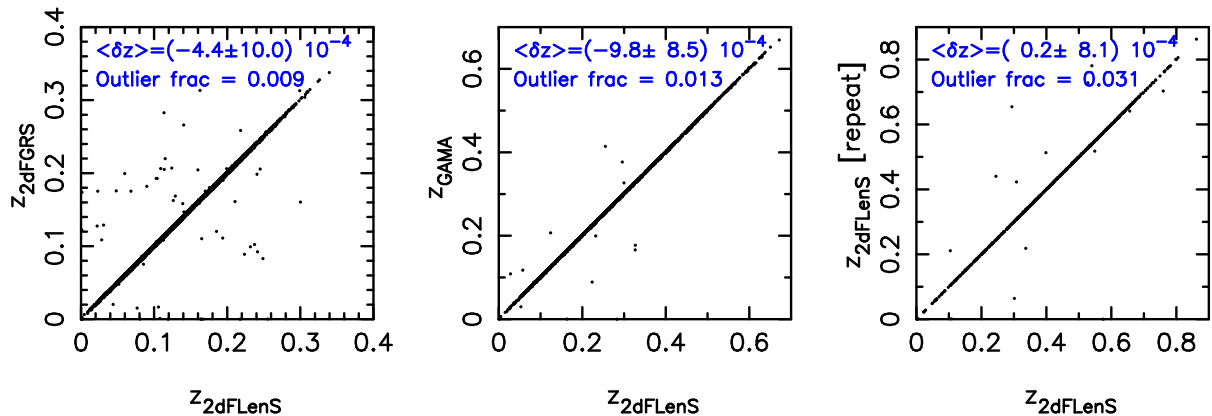}}}
\end{center}
\caption{Comparison of good-quality 2dFLenS redshifts with the
  redshifts of matched objects in the 2dFGRS (left-hand panel) and
  GAMA (middle panel) surveys, and the redshifts of repeated objects
  in 2dFLenS (right-hand panel).  The number of objects appearing in
  each plot are, from left-to-right, $6{,}384$, $3{,}224$ and $839$.}
\label{figredcheck}
\end{figure*}

The redshifts of 2dFLenS spectra may be determined using their
characteristic patterns of spectral lines in absorption and emission.
The incidence of spectral lines depends on the target type: for the
highest-priority LRGs, redshifts are typically derived from absorption
lines including Ca H (3935\AA) and K (3970\AA), H$\delta$ (4103\AA),
G-band (4304\AA), H$\beta$ (4863\AA), Mg (5169\AA) and Na (5895\AA).
Figure \ref{figstackspec} illustrates 2dFLenS LRG spectra stacked in
redshift slices of width $\Delta z = 0.1$.

We used a variety of tools to determine these redshifts.  Complete
automation of the redshifting process is problematic due to the noisy
nature of many of the spectra, and in particular the presence of
artefacts such as residuals from imperfect cosmic ray, sky removal and
splicing of the blue and red portions of the spectrum.  Therefore, all
spectra were visually inspected by 2dFLenS team members and assigned a
final integer quality flag $Q$ in the range 1-6.  These flag values
respectively indicate: unknown redshift ($Q=1$), a possible but
uncertain redshift ($Q=2$), a probably correct redshift derived from
noisy data or fewer spectral features ($Q=3$), a secure redshift
confirmed by multiple spectral features ($Q=4$), and a spectrum that
is clearly not extragalactic ($Q=6$).  The science analyses described
in this paper use $Q=3$ and $Q=4$ spectra.  The classification $Q=5$
is not used.

Three specific codes were used in the 2dFLenS redshifting process, two
of which include a visualization capability which team members used to
assign quality flags.

\begin{itemize}

\item {\sc runz} (Saunders, Cannon \& Sutherland 2004) is the AAO
  redshifting software with long-standing development spanning several
  AAT surveys such as 2dFGRS and WiggleZ.  {\sc runz} employs redshift
  determination from either discrete emission-line fitting or Fourier
  cross-correlation with a set of galaxy and stellar absorption-line
  templates (Tonry \& Davis 1979).  The {\sc runz} code may be
  executed without user interaction, but reliable assignment of
  redshift quality flags requires subsequent visual inspection of each
  spectrum.

\item {\sc autoz} (Baldry et al.\ 2014) is a fully-automatic
  cross-correlation redshifting code developed for the Galaxy And Mass
  Assembly (GAMA) survey.  In addition to a best-fitting redshift,
  {\sc autoz} also returns a figure-of-merit ${\rm cc}_{\rm fom}$
  which Baldry et al.\ (2014) relate to a quantitative confidence of
  redshift assignment.

\item {\sc marz} (Hinton et al.\ 2016) is an independent redshifting
  pipeline recently developed for the OzDES survey (Yuan et
  al.\ 2015).  {\sc marz} extends the matching algorithms of {\sc
    autoz} to include quasar templates, and offers a web-based
  visualization interface through which users can assign quality flags
  and manually redshift spectra as needed.

\end{itemize}

Two different processes were used by 2dFLenS team members for
assigning redshift quality flags to spectra.  First, all reduced
2dFLenS fields were passed through the {\sc autoz} code, and the
results were captured in an input file which may be visually inspected
using {\sc runz}.  The second possible process was to use {\sc marz}
for redshifting.

Possible redshifting errors and variations between 2dFLenS team
members in the optimism of redshift quality-flag assignment were
mitigated by subsequent inspection of borderline cases and potential
blunders.  In detail, all spectra flagged as bad-quality redshifts
with ${\rm cc}_{\rm fom} > 5$, or good-quality redshifts with ${\rm
  cc}_{\rm fom} < 3.5$, were checked for potential blunders.  Figure
\ref{figredcompcc} illustrates the relation between the fraction of
spectra assigned quality flags $Q \ge 3$, and the ${\rm cc}_{\rm fom}$
values assigned by the {\sc autoz} code.

In order to check the reliability of assigned redshifts, we compared
2dFLenS redshifts with external surveys (2dFGRS and GAMA) where
available, and also with repeat redshifts resulting from multiple
observations of a field.  Figure \ref{figredcheck} illustrates the
results of these comparisons.  Excluding outliers, the mean and
standard deviation of the quantities $z_{\rm 2dFLenS} - z_{\rm
  2dFGRS}$, $z_{\rm 2dFLenS} - z_{\rm GAMA}$ and $z_{\rm 2dFLenS, obs
  1} - z_{\rm 2dFLenS, obs 2}$ are $(-4.4 \pm 10.0) \times 10^{-4}$,
$(-9.8 \pm 8.5) \times 10^{-4}$ and $(0.2 \pm 8.1) \times 10^{-4}$,
respectively, consistent with zero difference in each case.  The
outlier fractions in these cases are $0.9\%$, $1.3\%$ and $3.1\%$,
respectively, which are negligible (and mostly consist of $Q=3$
spectra).

Redshifts are initially corrected to the heliocentric frame, and then
shifted to the rest-frame of the Cosmic Microwave Background (CMB)
radiation (Fixsen et al.\ 1996).  Our data catalogues and clustering
measurements are hence presented in the CMB frame.

\subsection{Survey status}
\label{secstatus}

\begin{figure*}
\begin{center}
\resizebox{12cm}{!}{\rotatebox{270}{\includegraphics{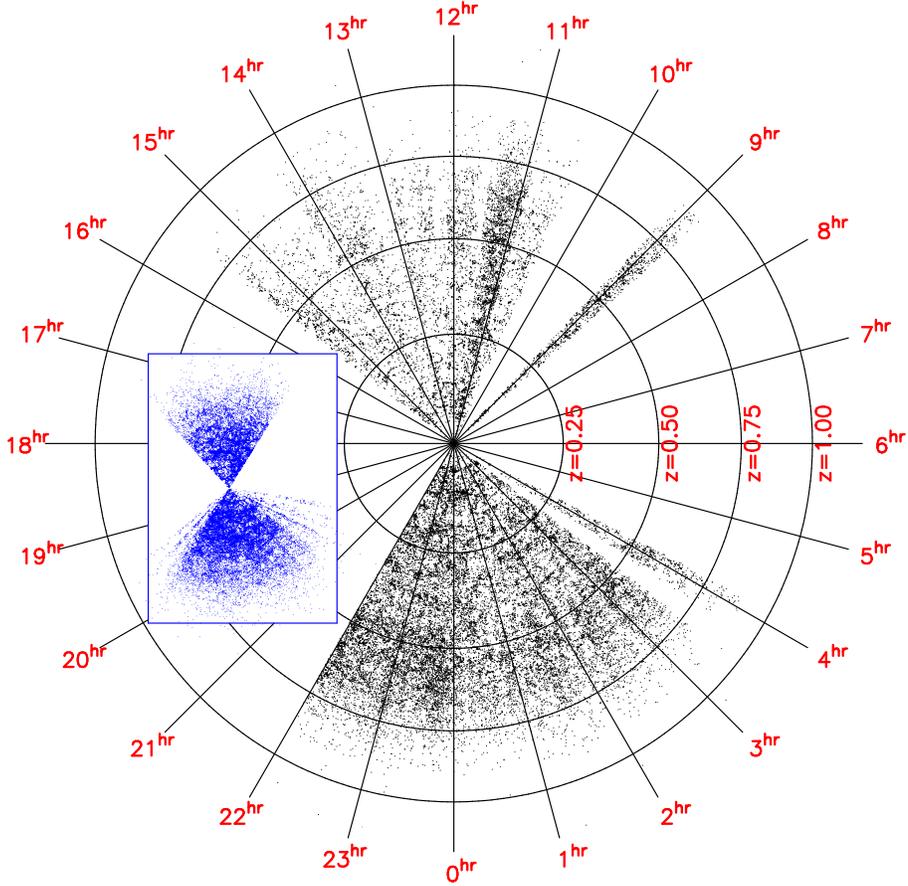}}}
\end{center}
\caption{Illustration of the distribution of large-scale structure in
  2dFLenS, generated by projecting the galaxy positions by right
  ascension and co-moving distance, indicated by the black points.
  The blue points in the inset display on the same scale the earlier
  2dFGRS dataset (Colless et al.\ 2001) obtained by the AAT.}
\label{figconeplot}
\end{figure*}

2dFLenS observations utilized $53.0$ allocated AAT nights and $3.0$
Director's nights between 17 Sep 2014 and 5 Jan 2016.  This
corresponded to a total of $465.5$ potential observing hours, of which
$293.6$ hrs ($63\%$) was clear, $161.8$ hrs ($35\%$) was lost to bad
weather and $10.1$ hrs ($2\%$) was lost to instrumentation fault.
During this time we observed 275 out of the 324 defined 2dFLenS AAT
pointing centres, with 18 additional re-observations due to poor
initial observing conditions.  Figure \ref{figprogress} illustrates
the final status of the survey coverage, mapping a total area of 731
deg$^2$.

These observations yielded a total of $70{,}079$ good-quality
redshifts, including $40{,}531$ LRGs and $28{,}269$ in the
magnitude-limited sample.  Table \ref{tabstats} lists the total number
of observed redshifts in each target class.  Figure \ref{figconeplot}
presents a projection of the positions of good-quality 2dFLenS
redshifts in comparison with earlier 2dFGRS observations in the same
field, illustrating the extension of the large-scale structure sample
to redshift $z=0.9$.

\begin{table*}
\caption{Number of targets observed, good redshifts obtained and the
  stellar fraction of those redshifts for each 2dFLenS sample.  The
  second row adds dual-use objects to the first row, that were
  selected for the magnitude-limited sample but flagged for
  observation in different target classes.  The lower half of the
  table lists the number of galaxies utilized in the clustering
  measurements described in Section \ref{secclus}, in redshift bins
  $0.15 < z < 0.43$ and $0.43 < z < 0.7$.}
\begin{center}
\begin{tabular}{cccc}
\hline
Target class & Spectra & Good redshifts & Stellar fraction \\
\hline
Complete mag-lim & $30{,}931$ & $28{,}269$ & $9\%$ \\
{\it (Including other classes)} & $(31{,}864)$ & $(29{,}123)$ & $(9\%)$ \\
Low-$z$ LRG & $15{,}004$ & $14{,}252$ & $2\%$ \\
Mid-$z$ LRG & $32{,}032$ & $19{,}376$ & $8\%$ \\
High-$z$ LRG & $18{,}116$ & $6{,}903$ & $6\%$ \\
Flux calibrator & $819$ & $819$ & $100\%$ \\
Spare fibre & $654$ & $460$ & $14\%$ \\
\hline
Total & $97{,}556$ & $70{,}079$ & $8\%$ \\
\hline
$0.15 < z < 0.43$ LRGs KiDS-S & & $8{,}473$ & \\
$0.15 < z < 0.43$ LRGs KiDS-N & & $3{,}556$ & \\
$0.43 < z < 0.7$ LRGs KiDS-S & & $13{,}402$ & \\
$0.43 < z < 0.7$ LRGs KiDS-N & & $4{,}036$ & \\
\hline
\end{tabular}
\end{center}
\label{tabstats}
\end{table*}

\begin{figure}
\begin{center}
\resizebox{8cm}{!}{\rotatebox{270}{\includegraphics{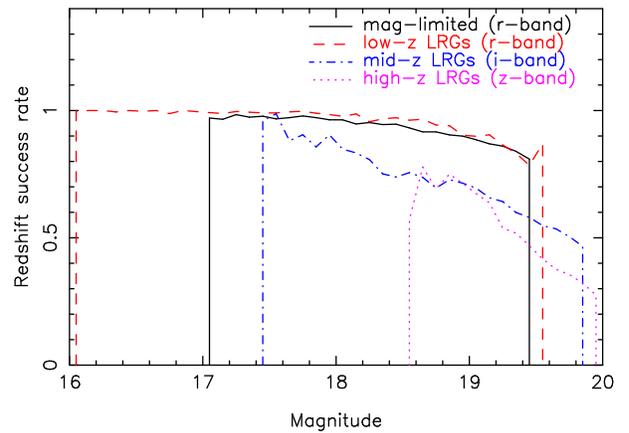}}}
\end{center}
\caption{Redshift success rate of 2dFLenS observations for each target
  class, as a function of the magnitude in the primary selection band
  in each case.  The black solid, red dashed, blue dot-dashed and
  magenta dotted lines show the magnitude-limited sample ($r$-band),
  the low-$z$ LRG sample ($r$-band), the mid-$z$ LRG sample ($i$-band)
  and high-$z$ LRG sample ($z$-band), respectively.}
\label{figredcompmag}
\end{figure}

The redshift completeness of 2dFLenS pointings (i.e., the fraction of
spectra with $Q \ge 3$) displays considerable variation between fields
driven primarily by the observing conditions (cloud cover and seeing),
and secondarily by airmass, with mean and standard deviation $71 \pm
15\%$.  Figure \ref{figredcompmag} displays the redshift success ($Q
\ge 3$) rate for each 2dFLenS target class, as a function of the
magnitude in the primary band for each selection.  The average
redshift success rate for the magnitude-limited, low-$z$ LRG, mid-$z$
LRG and high-$z$ LRG samples was $91\%$, $95\%$, $61\%$ and $38\%$,
respectively, with a gradual decline in the success rate with fainter
magnitudes.

\section{Selection function}
\label{secselfunc}

The selection function of a galaxy redshift survey describes the
variation in the expected mean number density of galaxies, at 3D
co-moving co-ordinate $\vec{r}$, in the absence of clustering.  An
accurate determination of the selection function is essential for
estimating the galaxy clustering statistics, which quantify
fluctuations relative to the mean density.  Our model for the
selection function of the 2dFLenS LRG samples considers angular
fluctuations in the density of the parent target catalogue on the sky,
the variation of the spectroscopic redshift completeness of each
AAOmega pointing with observing conditions, and the redshift
distribution of each target class together with its coupling to the
completeness.

We derived selection functions and performed clustering measurements
for two survey regions whose coverage is illustrated in Figure
\ref{figprogress}: KiDS-South (KiDS-S) and KiDS-North (KiDS-N).  The
KiDS-S analysis region is delineated by the boundaries listed in Table
\ref{tabregions}, and the KiDS-N region includes both the stripe of
2dFLenS pointings in the NGC area visible in Figure \ref{figprogress}
and the RCS1111 region, but excludes CFHTLS W1.

\subsection{Angular selection function}

\subsubsection{Parent target catalogue}
\label{secclussys}

\begin{figure*}
\begin{center}
\resizebox{12cm}{!}{\rotatebox{270}{\includegraphics{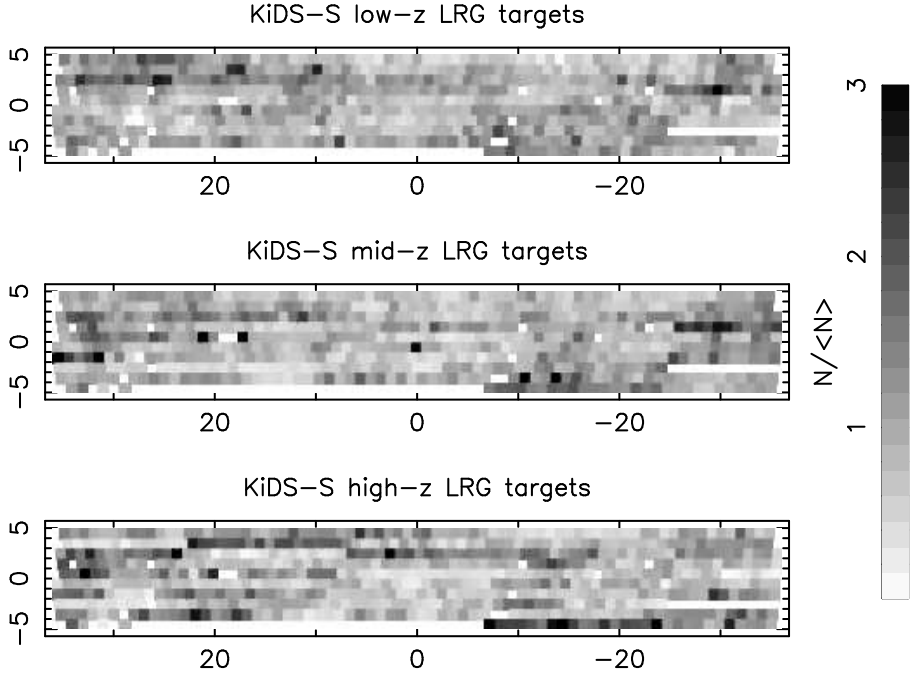}}}
\end{center}
\caption{Greyscale map showing the number of LRG targets selected in
  each ATLAS field in the KiDS-S region, relative to the mean.
  Significant field-to-field variations are apparent, resulting from
  photometric zero-point errors.  The $x$- and $y$-axes plot
  separation in degrees from the field centre.}
\label{figntarsys}
\end{figure*}

As described in Section \ref{secphotsys}, field-to-field variations in
the photometric accuracy of our ATLAS data reductions, of the order
$0.05-0.1$ magnitudes, imprint significant systematic fluctuations in
the number of selected LRG targets.  The situation is illustrated by
Figure \ref{figntarsys}, which displays the variation in the 1 deg
fields of the number density of the three 2dFLenS LRG samples in the
KiDS-S region.  If left uncorrected, these fluctuations would cause
significant systematic errors in the measured clustering.  Similar
effects are observed in the KiDS-N region.

In order to mitigate this effect we adopted a conservative approach to
the clustering analysis in which we marginalized over the unknown mean
(unclustered) number density in each field, constrained by the
observed number.  For each clustering statistic we generated an
ensemble of measurements corresponding to different angular selection
functions.  Each realization of the selection function was produced by
sampling the density $\lambda$ in each ATLAS field from a probability
distribution determined by the observed number of targets $N$ in that
field.  Assuming Poisson statistics the probability distribution is
given by
\begin{equation}
P(\lambda|N) \propto P(N|\lambda) \, P(\lambda) = \lambda^N \,
e^{-\lambda} / N! .
\end{equation}
For large $N$, this distribution is approximated by a Gaussian with
mean and variance equal to $N$.  We further increased the variance of
$P(\lambda)$ by adding in quadrature the contribution from angular
clustering.  The variance of galaxy counts-in-cells can be related to
an integral of the galaxy angular correlation function $w(\theta)$
over the cell area $A$ through
\begin{equation}
\langle (N - \langle N \rangle)^2 \rangle = \langle N \rangle +
\langle N \rangle^2 \frac{\int_{\rm cell} \int_{\rm cell} w(\theta) \,
  dA_1 \, dA_2}{A^2} .
\label{eqcountcell}
\end{equation}
We used the moments of BOSS galaxy counts in 1 deg$^2$ cells to
calibrate the final fraction in Equation \ref{eqcountcell}, which has
the value $\approx 0.04$, although varying this value does not
significantly affect our results.  The clustering contribution agrees
with that calculated from our own final $w(\theta)$ measurements
(using equations 1 and 2 in Blake \& Wall 2002).

\begin{figure*}
\begin{center}
\resizebox{14cm}{!}{\rotatebox{270}{\includegraphics{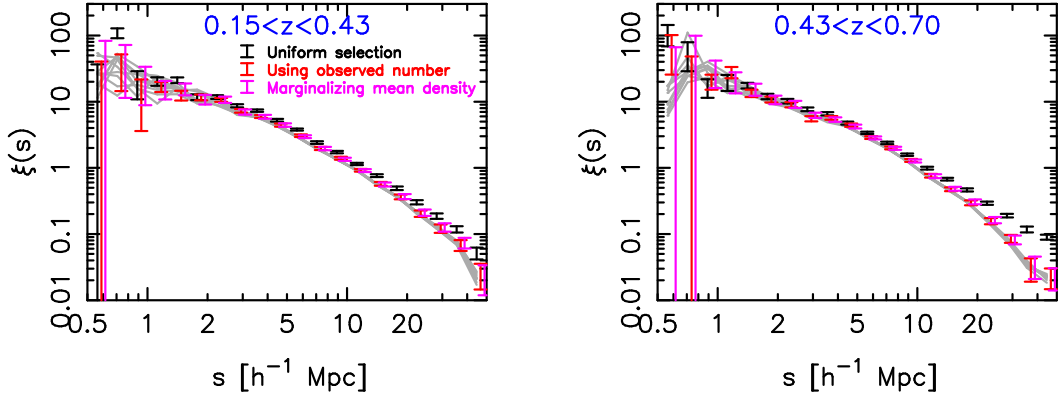}}}
\end{center}
\caption{Illustration of our clustering analysis marginalizing over
  the unknown mean galaxy density in each ATLAS field, for the case of
  the 2-point correlation function $\xi(s)$ in the KiDS-S region in
  the two redshift ranges $0.15 < z < 0.43$ (left panel) and $0.43 < z
  < 0.7$ (right panel).  The black data points are measurements
  assuming a uniform selection function, neglecting the systematic
  target density variations, and contain large-scale systematic
  errors.  These may be corrected using the observed number density in
  each tile as the angular selection, which produces the red data
  points.  The set of grey lines is an ensemble of clustering
  measurements in which the selection function is statistically
  sampled from a distribution defined by the observed number and
  clustering strength.  The magenta data points show the mean of this
  distribution, adding the density systematic error in quadrature to
  the original error.  Measurements for the different cases are
  slightly shifted along the $x$-axis for clarity.}
\label{figxisys}
\end{figure*}

With this ensemble of selection functions and derived clustering
measurements in place, we used their mean as our final determination
of each statistic, and added their covariance as a systematic error
contribution.  This process is illustrated by Figure \ref{figxisys}
for the case of the 2-point correlation function $\xi(s)$ in the
KiDS-S region.  Measurements assuming a uniform selection function,
neglecting the systematic variations apparent in Figure
\ref{figntarsys}, contain systematic error on large scales, which may
be corrected using the observed number density distribution as the
angular selection.  Marginalizing over the unknown mean density in
each field produces an ensemble of clustering measurements whose
variation defines a systematic error contribution.  {\it Importantly,
  we note that the magnitude of this systematic error is significantly
  less than the statistical error -- by typically an order of
  magnitude -- for all the clustering measurements considered in this
  paper.}  Therefore, whilst we always perform this marginalization
process, it does not have a significant impact on our results.

\subsubsection{Redshift completeness}

\begin{figure*}
\begin{center}
\resizebox{12cm}{!}{\rotatebox{270}{\includegraphics{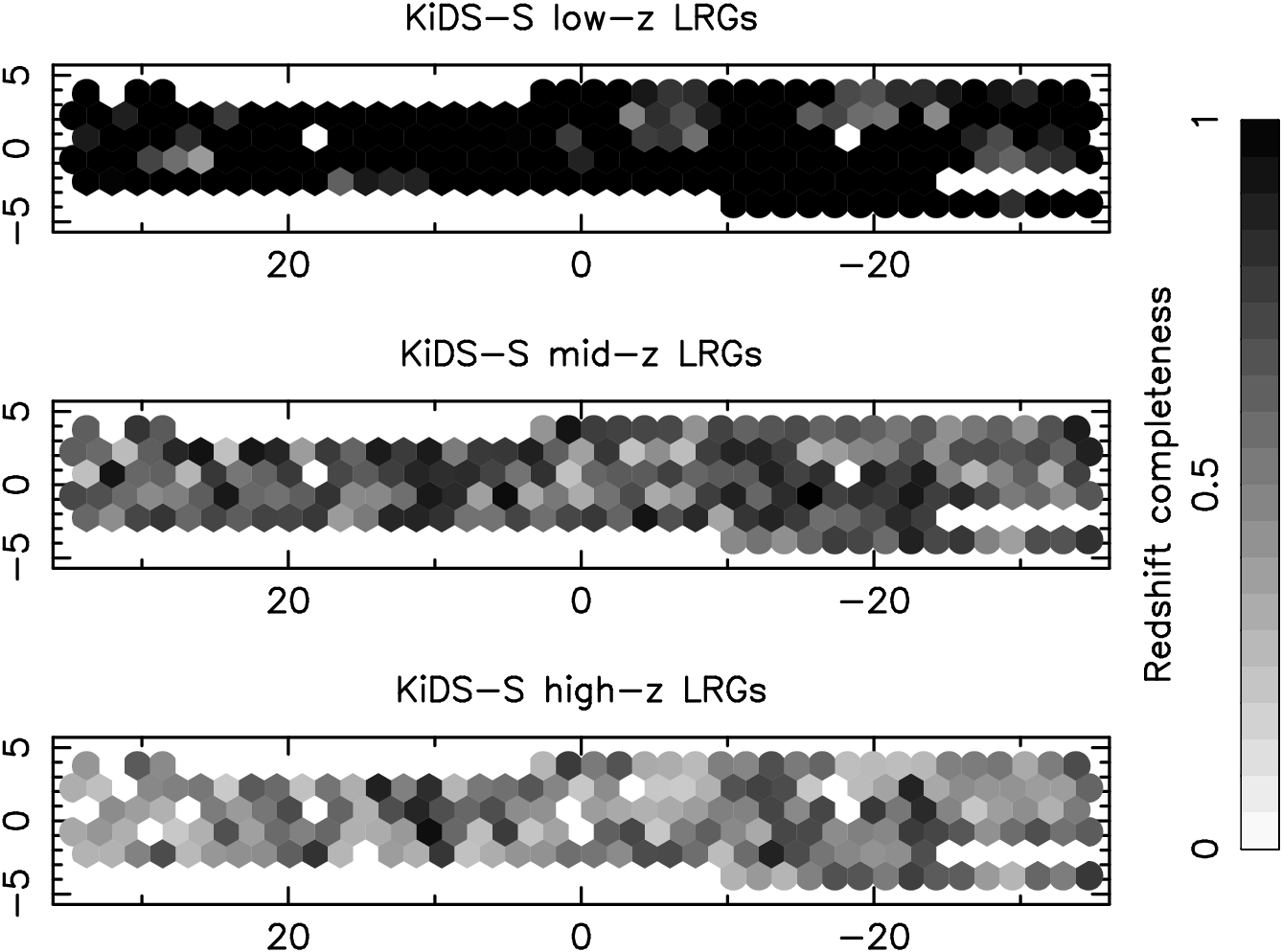}}}

\vspace{5mm}

\resizebox{12cm}{!}{\rotatebox{270}{\includegraphics{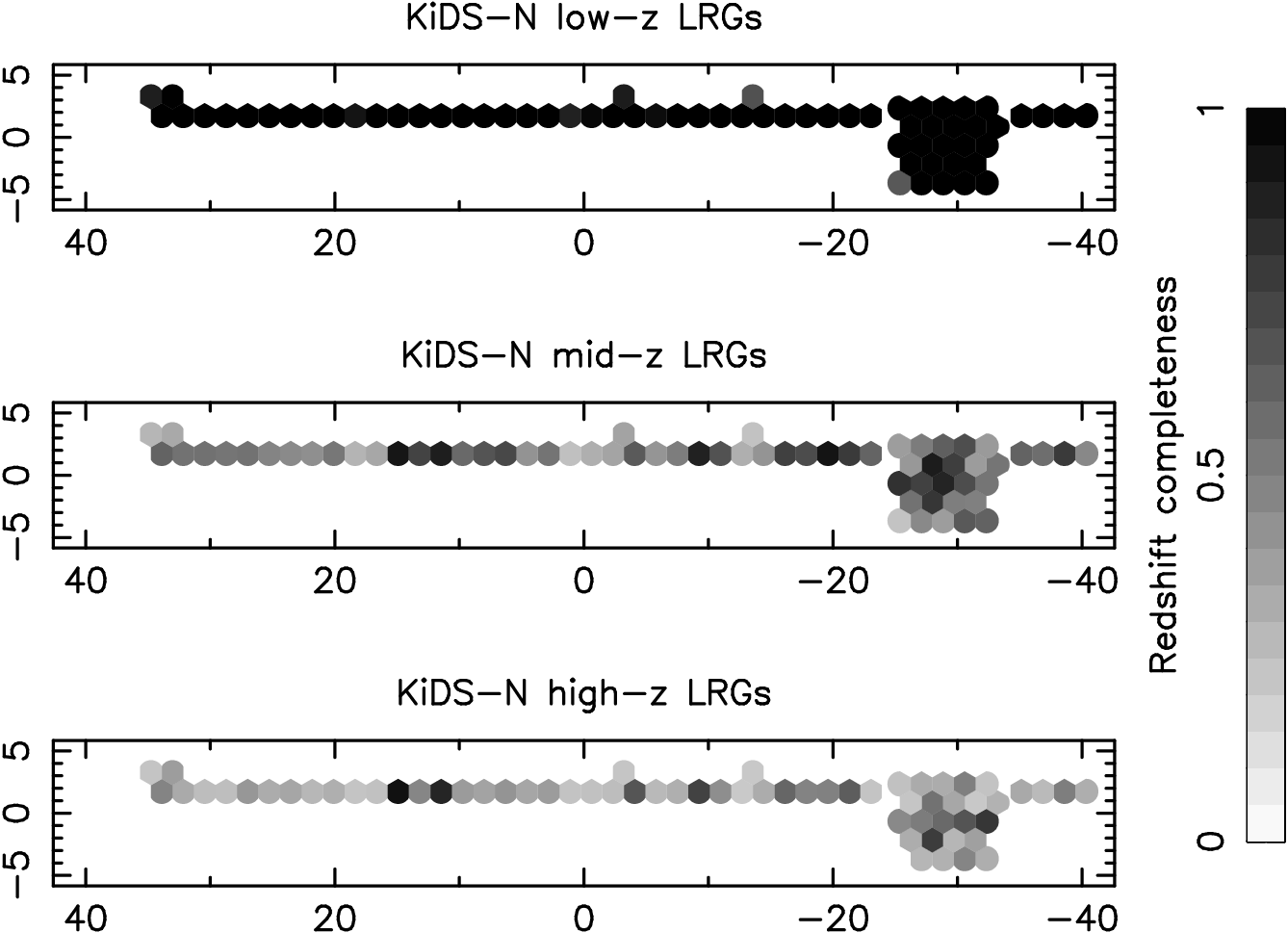}}}
\end{center}
\caption{Greyscale map showing the redshift completeness of 2dFLenS
  observations in the KiDS-S (top 3 rows) and KiDS-N (bottom 3 rows)
  regions.  The $x$- and $y$-axes plot separation in degrees from the
  field centre.}
\label{figredcompang}
\end{figure*}

LRG targets were uniquely assigned to the closest AAT field centre in
our pointing grid, producing a set of hexagonal survey
sectors\footnote{At the edges of the observing footprint, the sectors
  are bounded by circular arcs.}.  We modelled the variation in the
angular selection function due to incompleteness in redshift
determination using the ratio of good-quality redshifts to targets
within each of these sectors.  These redshift completeness maps are
displayed in Figure \ref{figredcompang} for the low-$z$, mid-$z$ and
high-$z$ LRG samples within the KiDS-S and KiDS-N survey regions.  The
low-$z$ LRG follow-up is highly complete, but the mid-$z$ and high-$z$
samples are imprinted with significant completeness variations driven
by AAT observing conditions.  We neglected any variation in the
redshift completeness across the 2-degree field-of-view (which may be
imprinted by either rotational mis-alignments in acquistion or
enhanced chromatic aberrations towards the edges of the field).  This
variation was lower than $\sim 10\%$ for 2dFLenS observations.

\subsection{Radial selection function}

We determined the redshift dependence of the selection function by
fitting parametric models (using Chebyshev polynomials) to the
empirical redshift distributions $N(z)$ of each LRG sample.  We
determined the order of the polynomial via a combination of
information criteria considerations and visual inspection.

\begin{figure}
\begin{center}
\resizebox{8cm}{!}{\rotatebox{270}{\includegraphics{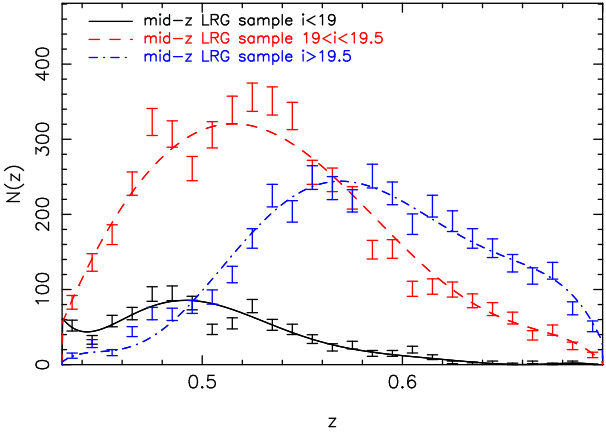}}}
\end{center}
\caption{The redshift distribution of the mid-$z$ LRG sample in
  $i$-band magnitude bands, together with Chebyshev polynomial fits.}
\label{fignzmag}
\end{figure}

Our model included the coupling between $N(z)$ and the angular
redshift completeness, such that our selection function is not
separable into angular and radial pieces.  This coupling arises
because in poorer observing conditions, corresponding to areas of
lower total redshift completeness, successful redshifts are
preferentially obtained for sources with brighter magnitudes (see
Figure \ref{figredcompmag}), which are preferentially located at lower
redshifts.  In detail, we fitted $N(z)$ functions for LRG samples in
bands of apparent magnitude, and constructed the model $N(z)$ within
each survey pointing using the magnitude distribution of galaxies with
successful redshifts within that pointing.  Figure \ref{fignzmag}
shows example $N(z)$ fits for the mid-$z$ LRG sample.

\subsection{Redshift bins}

The low-$z$, mid-$z$ and high-$z$ 2dFLenS LRG samples overlap in
redshift, as illustrated by Figure \ref{fignztype}.  We combined the
LRG samples into two independent redshift bins: $0.15 < z < 0.43$ and
$0.43 < z < 0.7$, weighting the selection function of each sample by
the relative target numbers.  The choice of these bins was motivated
by intended comparisons and combinations with the LOWZ and CMASS
samples of BOSS (Dawson et al.\ 2013), for example, to extend the
analysis of $E_G$ presented by Blake et al.\ (2016) in these redshift
bins.  Clustering measurements for 2dFLenS LRGs in these two redshift
bins will be presented in Section \ref{secclus}.

We computed the effective redshift of the selection functions in each
redshift bin as
\begin{equation}
z_{\rm eff} = \sum_{\vec{r}} z \, \left( \frac{n_g(\vec{r}) P_g}{1 +
  n_g(\vec{r}) P_g} \right)^2 ,
\end{equation}
where $n_g(\vec{r})$ is the mean galaxy number density in each grid
cell $\vec{r}$ and $P_g$ is the characteristic galaxy power spectrum
amplitude, which we evaluated at a scale $k = 0.1 \, h$ Mpc$^{-1}$
using the fiducial matter power spectrum and galaxy bias factors
specified in Section \ref{secclus}.  We obtained effective redshifts
$z_{\rm eff} = 0.31$ and $0.56$ in the two bins $0.15 < z < 0.43$ and
$0.43 < z < 0.7$.

\subsection{Fibre collisions}

\begin{figure}
\begin{center}
\resizebox{8cm}{!}{\rotatebox{270}{\includegraphics{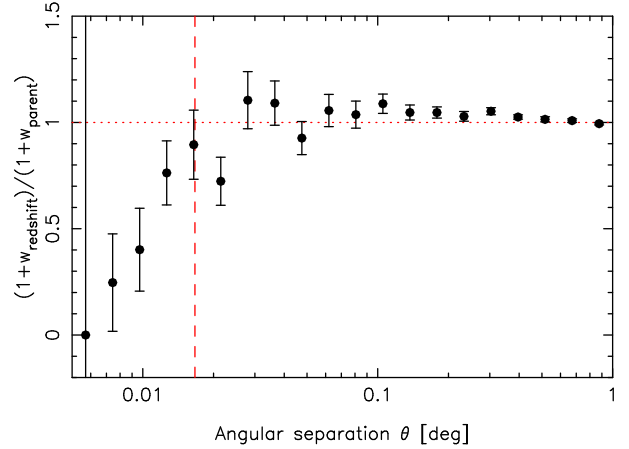}}}
\end{center}
\caption{The ratio of the angular correlation functions of the 2dFLenS
  parent and redshift catalogues, indicating the fraction of close
  pairs surviving the restrictions of fibre collisions as a function
  of angular separation.  Jack-knife errors are plotted.  The vertical
  dashed line corresponds to a separation of 1 arcmin, below which
  fibre collision effects are detectable.  The ratio is driven to
  values greater than 1 at moderate angular scales by stellar
  contamination in the parent catalogue, which reduces the value of
  $w_{\rm parent}$ compared to $w_{\rm redshift}$.}
\label{figfibcol}
\end{figure}

The minimum separation of the optical fibres of the 2dF spectrograph
is 30 arcsec, and there is a diminishing probability of observing in a
single pointing both members of a close pair of parent galaxies
separated by an angular distance of less than 2 arcmin.  This deficit
of close angular pairs in the redshift catalogue, known as `fibre
collisions', artificially suppresses the measured galaxy correlation
function on small scales.  We assess the deficit of close angular
pairs in Figure \ref{figfibcol} by plotting the ratio $(1 + w_z)/(1 +
w_p)$ as a function of angular separation $\theta$, where $w_z$ and
$w_p$ are the angular correlation functions of the redshift and parent
catalogues, respectively.  We measured the angular correlation
functions by applying the Landy-Szalay estimator (Landy \& Szalay
1993) to the positions of the data sources $D$ and a catalogue of
random sources $R$ which sample the survey selection function:
\begin{equation}
w(\theta) = \frac{DD(\theta) - 2 \, DR(\theta) + RR(\theta)}{RR(\theta)} .
\end{equation}
Significant effects are only detectable at the very smallest scales
$\theta < 1$ arcmin, and we do not correct for them in our analysis.

\section{Mock catalogues}
\label{secmock}

We determined the covariance of our 2dFLenS clustering statistics, and
their joint covariance with overlapping measurements of galaxy-galaxy
lensing and cosmic shear, using a set of mock catalogues created from
a large suite of N-body simulations which included a self-consistent
computation of gravitational lensing.

\subsection{SLICS catalogues}

Our mocks are built from the SLICS (Scinet LIght Cone Simulations)
series (Harnois-Deraps \& van Waerbeke 2015).  At the time of writing,
SLICS consisted of 930 N-body simulations created with the {\sc
  CUBEP$^3$M} code (Harnois-Deraps et al.\ 2013) using a WMAP9+BAO+SN
cosmological parameter set: matter density $\Omega_m = 0.2905$, baryon
density $\Omega_b = 0.0473$, Hubble parameter $h = 0.6898$, spectral
index $n_s = 0.969$ and normalization $\sigma_8 = 0.826$.  The
box-size of the simulations is $L = 505 \, h^{-1}$ Mpc, in which the
non-linear evolution of $1536^3$ particles is followed inside a
$3072^3$ grid cube.

For each simulation, the density field was output at 18 redshift
snapshots in the range $0 < z < 3$.  The gravitational lensing shear
and convergence is computed at these multiple lens planes using the
flat-sky Born approximation, and a survey cone spanning 60 deg$^2$ is
constructed by pasting together these snapshots.  In this process, the
planes were randomly shifted and the direction of the collapsed
dimension was changed in order to minimize residual correlations (see
Harnois-Deraps \& van Waerbeke 2015 for a complete description of the
light cone construction).  A spherical over-density halo finder was
executed on the particle data during the simulation run, producing
dark matter halo catalogues containing properties such as the mass,
position, center-of-mass velocity and 3-dimensional velocity
dispersion. These were then post-processed in order to select only
those that belonged to the light-cone geometry, self-consistently
reproducing the rotation and random shift imposed on the lens planes.

We used these simulation data products to build self-consistent mock
catalogues for overlapping cosmic shear and galaxy redshift surveys,
including realistic source and lens number densities, redshift
distributions and sampling of the density field.  We produced mocks
for two distinct cases.  First, we neglected the variation of the
angular selection function and generated 930 mocks, each of area 60
deg$^2$.  Secondly, we tiled together the individual simulations to
cover the area of our KiDS-N and KiDS-S regions (in a flat-sky
approximation).  The resulting tiled datasets could accommodate 65
mock catalogues using no simulated volume twice, which we sub-sampled
with the realistic angular selection functions of the cosmic shear and
galaxy redshift surveys.

The ensemble of 930 mocks is useful for determining the covariance of
a large data vector, such as an observation including cosmic shear
tomography, which can be area-scaled to match the true survey area.
The 65 larger mocks, which include the full selection function, permit
a more accurate determination of the covariance of the 2dFLenS
clustering measurements and were used in our analysis described in
Section \ref{secclus}.

\subsection{Halo occupation distribution}

We produced mock galaxy redshift survey catalogues by populating the
dark matter haloes of the N-body simulations using a Halo Occupation
Distribution (HOD) designed to match the measured large-scale
clustering amplitude of 2dFLenS galaxies.  For the purposes described
here, in which the small-scale `1-halo' clustering features are not
important and cannot be accurately measured due to fibre collisions
and low signal-to-noise ratio, we adopted a central galaxy HOD such
that the probability that a dark matter halo of mass $M$ hosts an LRG
transitions from 0 to 1 according to
\begin{equation}
P(M) = \frac{1}{2} \left[ 1 + {\rm erf} \left( \frac{\log_{10} M -
    \log_{10} M_0}{\sigma_{\log{M}}} \right) \right] ,
\label{eqhod}
\end{equation}
where $M_0$ and $\sigma_{\log{M}}$ are free parameters, and we
neglected satellite galaxies.  After populating dark matter haloes in
this manner, placing the mock galaxy at the central position of the
halo and assigning it the halo's centre-of-mass velocity, we
sub-sampled the mock galaxy distribution to match the 3D selection
function of the 2dFLenS galaxies in each survey region, deriving a
redshift-space position.  We varied the parameters $M_0$ and
$\sigma_{\log{M}}$ to reproduce the measured 2dFLenS clustering,
finding that an acceptable match was produced by the choices $M_0 =
10^{14.1} \, h^{-1} M_\odot$ and $\sigma_{\log{M}} = 0.2$,
i.e.\ cluster-scale halos.  This value of $M_0$ falls at the upper end
of the range found in fits to BOSS-CMASS LRGs (Guo et al.\ 2014),
consistent with the lower number density of 2dFLenS LRGs and our
neglect of the satellite contribution.

\begin{figure*}
\begin{center}
\resizebox{14cm}{!}{\rotatebox{270}{\includegraphics{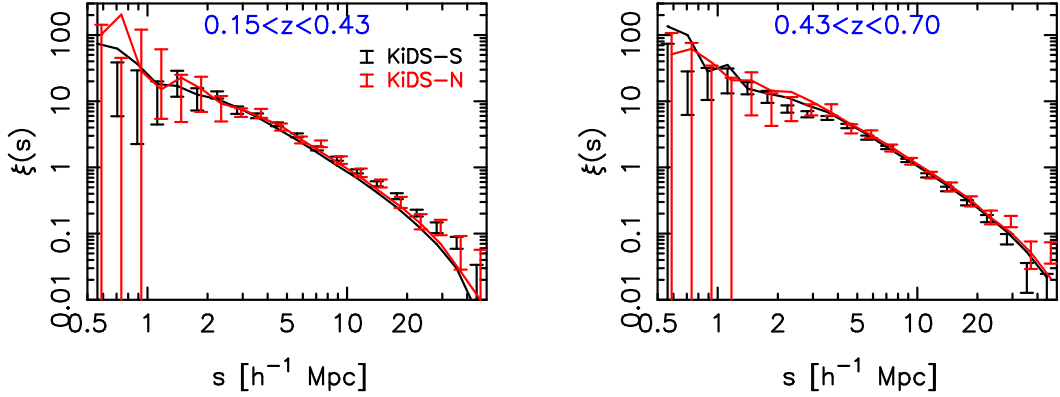}}}
\end{center}
\caption{The spatial correlation function $\xi_0(s)$ measured in the
  KiDS-S and KiDS-N survey regions (black and red data points)
  compared to similar measurements in the mock catalogues (solid
  lines).}
\label{figxidatamock}
\end{figure*}

We note that the systematic photometric variations described in
Section \ref{secphotsys} produced an artificial increase in the survey
selection function in some (small) areas, resulting in a target galaxy
number density which cannot be matched by haloes selected via Equation
\ref{eqhod}.  In these areas, we supplemented the mock catalogue by
randomly sampling haloes with masses $M > 10^{13.8} \, h^{-1}
M_\odot$, until the target number density was matched.  A comparison
of the spatial correlation functions $\xi_0(s)$ of the data and mock
catalogues is shown in Figure \ref{figxidatamock}, illustrating the
agreement produced by our approach.

\subsection{Joint lensing catalogues}

For science analyses requiring joint lensing and clustering mocks, we
produced the mock lensing catalogues using the approach described by
Joudaki et al.\ (2016a), which we briefly summarize here.
\begin{itemize}
\item We populated each simulation cone using a source redshift
  distribution and an effective source density matching that of the
  lensing survey, by Monte-Carlo sampling sources from the simulation
  density field.
\item We assigned two-component gravitational shears $(\gamma_1,
  \gamma_2)$ to each mock source by linearly interpolating the shear
  fields at the source positions between the values at adjacent
  snapshot redshifts.
\item We applied shape noise to the mock sources, drawing the noise
  components from a Gaussian distribution with standard deviation
  matching that of the lensing survey.
\end{itemize}
We note that, although sources in the cosmic shear survey dataset have
optimal weights determined by the shape measurement process, we
produced lensing mocks in which all sources have uniform weight, and
the varying weights are absorbed into the effective source density,
redshift distribution and shape noise.

\section{Clustering measurements}
\label{secclus}

In this Section we present clustering measurements of 2dFLenS LRGs
using three statistics.  First, we determined the projected
correlation function $w_p(r_p)$ as a function of transverse separation
$r_p$, in which the effect of redshift-space distortion is removed by
integrating along the line-of-sight direction (Section
\ref{secprojcorr}).  The projected correlation function is used to
estimate the bias of the galaxy sample and is also required for
determining the gravitational slip statistic $E_G$ as a test of
gravitational physics (Amon et al.\ 2016).  We also computed two
statistics which quantify the dependence of the clustering amplitude
on the angle to the line-of-sight: the correlation function multipoles
$\xi_\ell(s)$ (Section \ref{secximult}) and the power spectrum
multipoles $P_\ell(k)$ (Section \ref{secpkmult}).  These statistics
are used to fit models for redshift-space distortion, and we also test
whether consistent results are produced in Fourier space and
configuration space.  We converted the galaxy angular positions and
redshifts into 3D co-moving space using a flat $\Lambda$CDM fiducial
cosmology with matter density $\Omega_{\rm m} = 0.3$.

\subsection{Projected correlation function and galaxy bias}
\label{secprojcorr}

We estimated the projected correlation function of 2dFLenS galaxies by
initially measuring the 2D correlation function $\xi(r_p,\Pi)$ as a
function of projected pair separation $r_p$ and line-of-sight
separation $\Pi$ using the Landy-Szalay estimator:
\begin{equation}
\xi(r_p,\Pi) = \frac{DD(r_p,\Pi) - 2 \, DR(r_p,\Pi) +
  RR(r_p,\Pi)}{RR(r_p,\Pi)} .
\label{eqxiest}
\end{equation}
In each 2dFLenS survey region we generated a random catalogue 10 times
larger than the data catalogue.  In Equation \ref{eqxiest}, $DD$, $DR$
and $RR$ are the data-data, data-random and random-random pair counts
in each separation bin.  For a pair of galaxies with position vectors
$\vec{r}_1$ and $\vec{r}_2$, mean position $\vec{r} = (\vec{r}_1 +
\vec{r}_2)/2$ and separation vector $\vec{s} = \vec{r_2} - \vec{r_1}$,
the separation bin values are defined by $\Pi =
|\vec{s}.\vec{r}|/|\vec{r}|$ and $r_p = \sqrt{|\vec{s}|^2 - \Pi^2}$.

\begin{figure*}
\begin{center}
\resizebox{14cm}{!}{\rotatebox{270}{\includegraphics{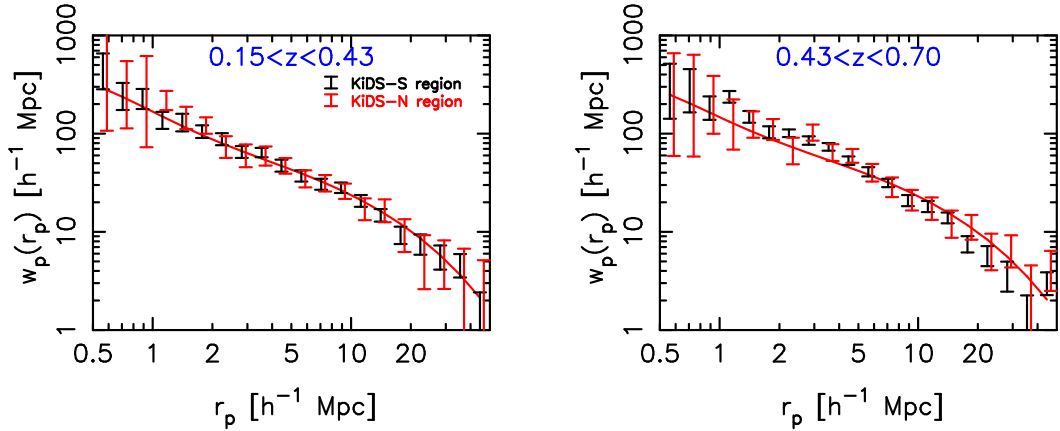}}}
\end{center}
\caption{The projected correlation function $w_p(r_p)$ for 2dFLenS
  LRGs in the KiDS-S and KiDS-N regions.  Results are shown for two
  redshift ranges $0.15<z<0.43$ (left) and $0.43<z<0.7$ (right) with
  errors estimated using the ensemble of mock catalogues.  The solid
  lines indicate the best fit of a single-parameter bias model in our
  fiducial cosmology.}
\label{figwp}
\end{figure*}

We then determined the projected correlation function using the sum
\begin{equation}
w_p(r_p) = 2 \sum_i \xi(r_p, \Pi_i) \, \Delta \Pi_i ,
\end{equation}
where we summed over 10 logarithmically-spaced bins in $\Pi$ from
$0.1$ to $60 \, h^{-1}$ Mpc.  The measurements of $w_p(r_p)$ for
2dFLenS galaxies in 20 logarithmically-spaced bins in $r_p$ from $0.5$
to $50 \, h^{-1}$ Mpc are shown in Figure \ref{figwp}, for the two
redshift ranges $0.15 < z < 0.43$ and $0.43 < z < 0.7$.  Errors are
obtained from the mock catalogues.

For illustrative purposes we show the fit of a single-parameter bias
model to the data using a non-linear power spectrum $P_{\rm m}(k)$
computed in a fiducial cosmology.  We generated $P_{\rm m}(k)$ using
the non-linear corrections calibrated by Takahashi et al.\ (2012) as
implemented by the {\sc camb} software package (Lewis, Challinor \&
Lasenby 2000).  For the purposes of this measurement we specified the
cosmological parameters used to generate the model power spectrum as
the maximum-likelihood (``TT+lowP'') parameters fit to {\sl Planck}
CMB observations and quoted in the 1st column of Table 3 in Planck
collaboration (2015): physical baryon density $\Omega_{\rm b} h^2 =
0.02222$, physical cold dark matter density $\Omega_{\rm c} h^2 =
0.1197$, Hubble parameter $H_0 = 67.31$ km s$^{-1}$ Mpc$^{-1}$,
spectral index $n_{\rm s} = 0.9655$ and normalization $\sigma_8 =
0.829$.\footnote{There is a minor inconsistency between the inferred
  value of $\Omega_{\rm m} = 0.315$ and that assumed for the fiducial
  survey geometry, $\Omega_{\rm m} = 0.3$, although the two values are
  statistically consistent given the error in the {\sl Planck}
  measurement, and the corresponding Alcock-Paczynski distortion is
  negligible.}  The best-fitting bias parameters are $b = 1.84 \pm
0.03$ ($0.15 < z < 0.43$) and $b = 2.10 \pm 0.03$ ($0.43 < z < 0.7$).

\subsection{Multipole correlation functions}
\label{secximult}

We estimated the redshift-space correlation function $\xi(s,\mu)$ as a
function of co-moving separation $s = |\vec{s}|$ and the cosine of the
angle of the pair separation vector with respect to the line-of-sight
towards the mean position $\vec{r}$, $\mu = |\vec{s}.\vec{r}| /
|\vec{s}| |\vec{r}|$, using a Landy-Szalay estimator equivalent to
Equation \ref{eqxiest}.  For this estimate we assigned each galaxy
optimal `FKP' weights (Feldman, Kaiser \& Peacock 1994) defined by
\begin{equation}
w_{\rm FKP}(\vec{r}) = \frac{1}{1 + n_g(\vec{r}) \, P_g} ,
\end{equation}
where $n_g(\vec{r})$ is the galaxy number density at position
$\vec{r}$ expected in the mean realization of the survey selection
function, and $P_g$ is a characteristic value of the power spectrum,
which we take as $P_g = 20{,}000 \, h^{-3}$ Mpc$^3$ motivated by the
power spectrum measurements presented in Section \ref{secpkmult}.
This weighting scheme ensures equal weight per volume where the
measurement is limited by sample variance ($n_g P_g \gg 1$) and equal
weight per galaxy where the measurement is limited by shot noise ($n_g
P_g \ll 1$).  We used 9 separation bins of width $\Delta s = 10 \,
h^{-1}$ Mpc in the range $s < 90 \, h^{-1}$ Mpc, and 100 angular bins
of width $\Delta \mu = 0.01$.

It is convenient to compress the information encoded in $\xi(s,\mu)$
into correlation function multipoles defined by
\begin{equation}
\xi_\ell(s) = \frac{2 \ell + 1}{2} \int_{-1}^1 d\mu \, \xi(s,\mu) \,
L_\ell(\mu) ,
\label{eqximultdef}
\end{equation}
where $L_\ell$ is the Legendre polynomial of order $\ell$.  The
linear-theory contribution to the clustering is described by a
summation over terms $\ell = \lbrace 0, 2, 4 \rbrace$.  The monopole
$\xi_0(s)$ represents the total angle-averaged spatial correlation
function; the quadrupole $\xi_2(s)$ encodes the leading-order
redshift-space distortion signal.  We estimated $\xi_\ell(s)$ in
Equation \ref{eqximultdef} by converting the integral into a sum over
$\mu$-bins.

\begin{figure*}
\begin{center}
\resizebox{14cm}{!}{\rotatebox{270}{\includegraphics{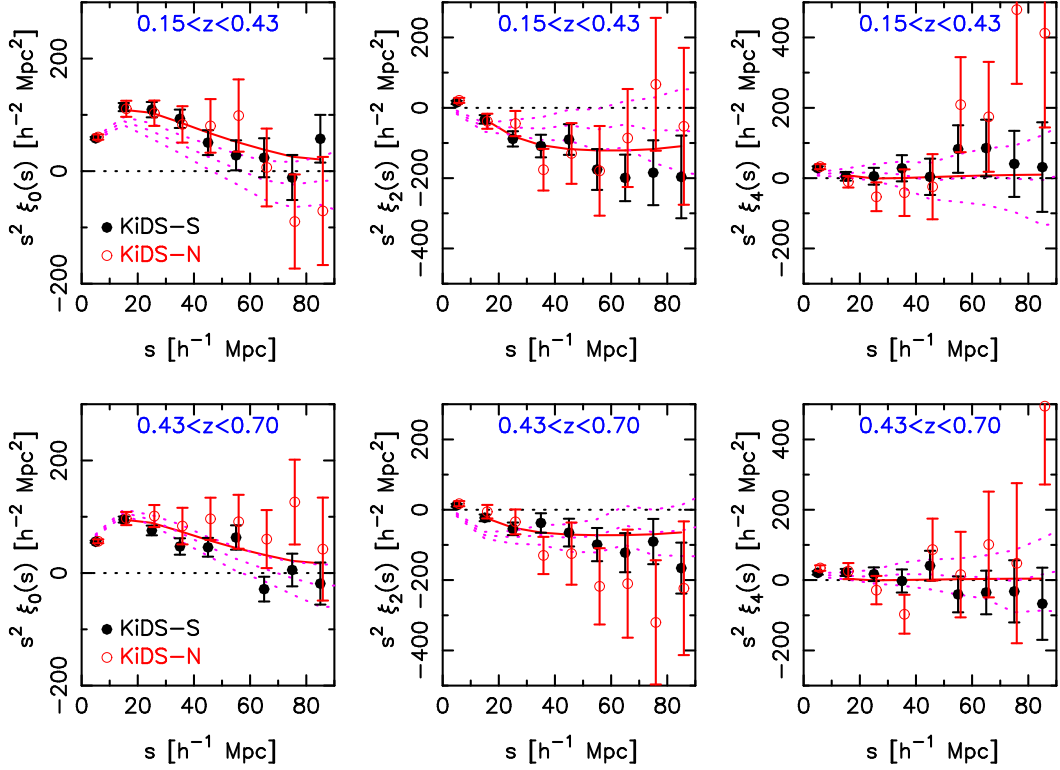}}}
\end{center}
\caption{The multipole correlation functions $(\xi_0, \xi_2, \xi_4)$,
  from left-to-right, for 2dFLenS LRGs in the KiDS-S and KiDS-N
  regions.  Results are shown for two redshift ranges $0.15<z<0.43$
  (top row) and $0.43<z<0.7$ (bottom row), and scaled by $s^2$ for
  clarity of presentation.  The solid red line indicates the
  best-fitting model and the magenta dotted lines display the mock
  mean and $68\%$ confidence range of mock measurements for the KiDS-S
  region.}
\label{figximultdata}
\end{figure*}

Figure \ref{figximultdata} displays our measurement of the multipole
correlation functions for 2dFLenS data in the KiDS-S and KiDS-N
regions for redshift ranges $0.15 < z < 0.43$ and $0.43 < z < 0.7$.
We are able to detect the signature of redshift-space distortion via
the non-zero values of the quadrupole $\xi_2(s)$; the hexadecapole
$\xi_4(s)$ is consistent with zero.  We overplot the best-fitting
redshift-space distortion model (see Section \ref{secrsd}).

\begin{figure*}
\begin{center}
\resizebox{12cm}{!}{\rotatebox{270}{\includegraphics{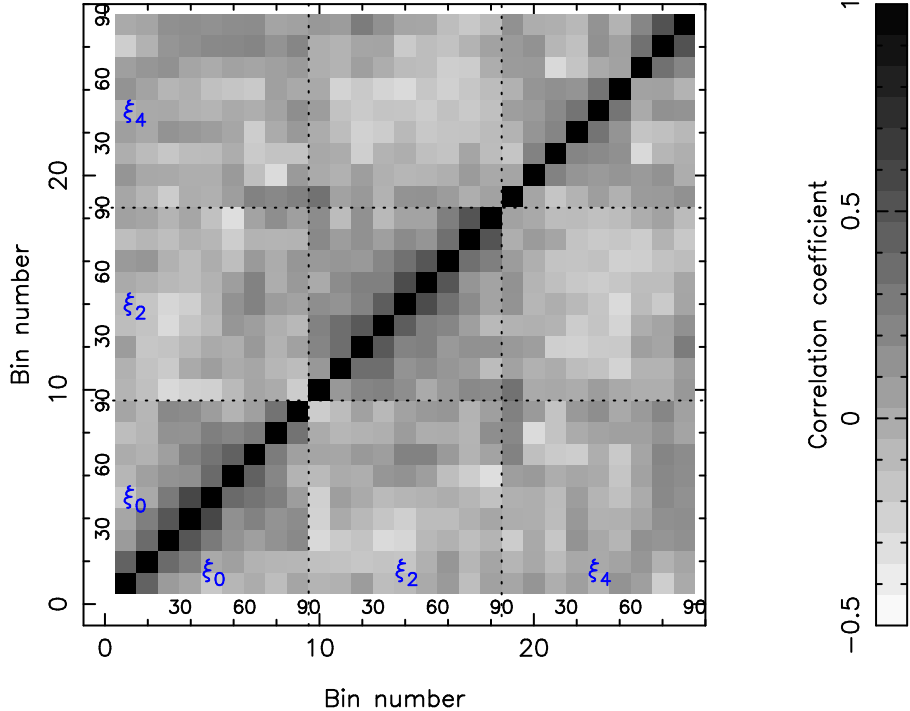}}}
\end{center}
\caption{The correlation matrix for the 2dFLenS multipole correlation
  functions arranged in a data vector $\xi_{\rm est}(i) = \lbrace
  \xi_0, \xi_2, \xi_4 \rbrace$, derived from the mock catalogues as
  ${\rm Cov}_{ij}/\sqrt{ {\rm Cov}_{ii} \, {\rm Cov}_{jj}}$.  Results
  are shown for the redshift range $0.43<z<0.7$ for the KiDS-S region;
  they are similar for the redshift range $0.15<z<0.43$ and for
  KiDS-N.  The labels $(30, 60, 90)$ denote separations in $h^{-1}$
  Mpc.}
\label{figximultcov}
\end{figure*}

We estimated the correlation function multipoles for each of the mock
catalogues, and used the measurements for the ensemble of realizations
to construct a covariance matrix
\begin{equation}
{\rm Cov}_{ij} = \langle \xi_{\rm est}(i) \, \xi_{\rm est}(j) \rangle
- \langle \xi_{\rm est}(i) \rangle \, \langle \xi_{\rm est}(j) \rangle ,
\end{equation}
where the array $\xi_{\rm est}(i)$ consists of the concatenation
\begin{equation}
\lbrace \xi_0(s_1), \xi_0(s_2), ..., \xi_2(s_1), \xi_2(s_2), ...,
\xi_4(s_1), \xi_4(s_2), ... \rbrace .
\end{equation}
The corresponding correlation matrix, defined by ${\rm
  Cov}_{ij}/\sqrt{{\rm Cov}_{ii} \, {\rm Cov}_{jj}}$, is displayed in
Figure \ref{figximultcov} for the KiDS-S region for the redshift range
$0.43 < z < 0.7$ (results for the other region and redshift range are
similar).  The off-diagonal correlations are typically low, and
dominated by neighbouring separation bins.  We also plot in Figure
\ref{figximultdata} the $68\%$ confidence range of the mock
correlation function measurements, within which the data points
generally lie.

\subsection{Multipole power spectra}
\label{secpkmult}

The dependence of the galaxy clustering amplitude on the angle to the
line-of-sight, including redshift-space distortion, may be quantified
in Fourier space using multipole power spectra $P_\ell(k)$:
\begin{equation}
P(k,\mu) = \sum_\ell P_\ell(k) \, L_\ell(\mu) .
\label{eqpkmultdef1}
\end{equation}
The orthogonality of $L_\ell(\mu)$ implies that
\begin{equation}
P_\ell(k) = \frac{2 \ell + 1}{2} \int_{-1}^1 d\mu \, P(k, \mu) \,
L_\ell(\mu) .
\label{eqpkmultdef2}
\end{equation}

\subsubsection{Power spectrum estimation}
\label{secpkmultest}

We estimated the multipole power spectra $\lbrace P_0(k), P_2(k),
P_4(k) \rbrace$ using the direct Fast Fourier Transform (FFT) method
presented by Bianchi et al.\ (2015) and Scoccimarro (2015).  The use
of FFTs results in a significant speed-up compared to the estimation
by direct summation described earlier by Yamamoto et al.\ (2006),
Blake et al.\ (2011a) and Beutler et al.\ (2014).

We first enclosed the survey cone within a cuboid of sides $(L_x, L_y,
L_z)$ and gridded the catalogue of $N$ galaxies in cells numbering
$(n_x, n_y, n_z)$ using nearest grid point assignment to produce a
distribution $n(\vec{r})$, where $\sum_{\vec{r}} n(\vec{r}) = N$.  The
cell dimensions were chosen such that the Nyquist frequencies in each
direction (e.g.\ $k_{\rm Nyq} = \pi n_x/L_x$) exceeded the maximum
frequency of measured power by at least a factor of 3.  We then
defined the weighted overdensity field
\begin{equation}
F(\vec{r}) = w_{\rm FKP}(\vec{r}) \, \left[ n(\vec{r}) - N \,
  W(\vec{r}) \right] ,
\end{equation}
where $W(\vec{r})$ is proportional to the survey selection function
determined in Section \ref{secselfunc}, which describes the number of
galaxies expected in each cell $\vec{r}$ in the absence of clustering
assuming the normalization $\sum_{\vec{r}} W(\vec{r}) = 1$.

We employed the following estimators for the power spectrum multipoles
(Bianchi et al.\ 2015, equations 6-8):
\begin{eqnarray}
& & P_0(\vec{k}) = \frac{1}{I} \, A_0(\vec{k}) \, A_0^*(\vec{k}) - P_{\rm noise} , \\
& & P_2(\vec{k}) = \frac{5}{2I} \, A_0(\vec{k}) \, \left[ 3 A_2^*(\vec{k}) - A_0^*(\vec{k}) \right] , \\
& & P_4(\vec{k}) = \frac{9}{8I} \, A_0(\vec{k}) \, \left[ 35 A_4^*(\vec{k}) - 30 A_2^*(\vec{k}) + 3 A_0^*(\vec{k}) \right] ,
\end{eqnarray}
in terms of the variables
\begin{eqnarray}
& & A_n(\vec{k}) = \int d^3\vec{r} \, ( \hat{\vec{k}}.\hat{\vec{r}} )^n \, F(\vec{r}) \, \exp{(i \vec{k}.\vec{r})} , \\
& & P_{\rm noise} = \int d^3\vec{r} \, w_{\rm FKP}(\vec{r})^2 \, n(\vec{r}) , \\
& & I = N^2 \int d^3\vec{r} \, w_{\rm FKP}(\vec{r})^2 \, W(\vec{r})^2 .
\end{eqnarray}
We determined the functions $A_n(\vec{k})$ by evaluating the following
quantities using FFTs:
\begin{eqnarray}
& & A_0(\vec{k}) = \int d^3\vec{r} \, F(\vec{r}) \, \exp{(i \vec{k}.\vec{r})} , \\
& & B_{ij}(\vec{k}) = \int d^3\vec{r} \, b_{ij}(\vec{r}) \, F(\vec{r}) \, \exp{(i \vec{k}.\vec{r})} , \\
& & C_{ijk}(\vec{k}) = \int d^3\vec{r} \, c_{ijk}(\vec{r}) \, F(\vec{r}) \, \exp{(i \vec{k}.\vec{r})} ,
\end{eqnarray}
where
\begin{eqnarray}
& & b_{ij}(\vec{r}) = \frac{r_i \, r_j}{r^2} , \\
& & c_{ijk}(\vec{r}) = \frac{r_i^2 \, r_j \, r_k}{r^4} .
\end{eqnarray}
The indices $(i,j,k)$ range over $\lbrace 1,2,3 \rbrace$, where $(r_1,
r_2, r_3) = (x, y, z)$.  In terms of these variables,
\begin{eqnarray}
A_2(\vec{k}) &=& \sum_{ij} \beta_{ij}(\vec{k}) \, B_{ij}(\vec{k}) \\
&=& \frac{1}{k^2} \lbrace k_x^2 B_{xx} + k_y^2 B_{yy} + k_z^2 B_{zz} \nonumber \\
& & + 2 \left[ k_x k_y B_{xy} + k_x k_z B_{xz} + k_y k_z B_{yz} \right] \rbrace ,
\end{eqnarray}
and
\begin{eqnarray}
A_4(\vec{k}) &=& \sum_{ijk} \gamma_{ijk}(\vec{k}) \, C_{ijk}(\vec{k}) \\
&=& \frac{1}{k^4} \lbrace k_x^4 C_{xxx} + k_y^4 C_{yyy} + k_z^4 C_{zzz} \nonumber \\
& & + 4 \left[ k_x^3 k_y C_{xxy} + k_x^3 k_z C_{zzx} + k_y^3 k_x C_{yyx} \right] \nonumber \\
& & + 4 \left[ k_y^3 k_z C_{yyz} + k_z^3 k_x C_{zzx} + k_z^3 k_y C_{zzy} \right] \nonumber \\
& & + 6 \left[ k_x^2 k_y^2 C_{xyy} + k_x^2 k_z^2 C_{xzz} + k_y^2 k_z^2 C_{yzz} \right] \nonumber \\
& & + 12 \, k_x k_y k_z \left[ k_x C_{xyz} + k_y C_{yxz} + k_z C_{zxy} \right] \rbrace .
\end{eqnarray}
We obtained the final power spectrum multipoles $\lbrace P_0(k),
P_2(k), P_4(k) \rbrace$ by angle-averaging $P_\ell(\vec{k})$ in
spherical shells in $\vec{k}$-space.

\begin{figure*}
\begin{center}
\resizebox{14cm}{!}{\rotatebox{270}{\includegraphics{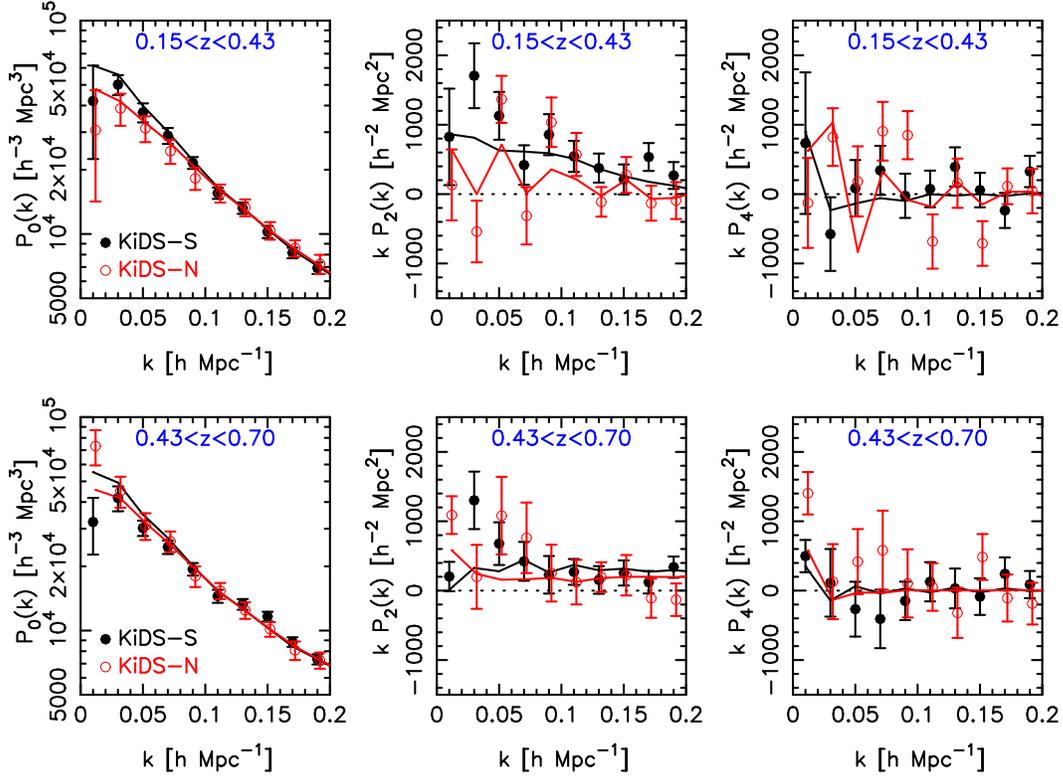}}}
\end{center}
\caption{The multipole power spectra $(P_0, P_2, P_4)$, from
  left-to-right, for 2dFLenS LRGs in the KiDS-S (black solid circles)
  and KiDS-N (red open circles) regions.  Results are shown for two
  redshift ranges $0.15<z<0.43$ (top row) and $0.43<z<0.7$ (bottom
  row).  The solid lines indicate the best-fitting model convolved in
  each case with the region window function, which produces the
  `choppy' appearance in the model.}
\label{figpkmultdata}
\end{figure*}

Our measurements of the multipole power spectra $\lbrace P_0, P_2, P_4
\rbrace$ for the 2dFLenS KiDS-S and KiDS-N regions, for the two
redshift ranges $0.15 < z < 0.43$ and $0.43 < z < 0.7$, are displayed
in Figure \ref{figpkmultdata}.  We used 10 Fourier bins of width
$\Delta k = 0.02 \, h$ Mpc$^{-1}$ in the range $0 < k < 0.2 \, h$
Mpc$^{-1}$.  A clear detection of non-zero quadrupole $P_2(k)$ is
again obtained, and the hexadecapole $P_4(k)$ is consistent with zero.
We overplot the best-fitting redshift-space distortion model (see
Section \ref{secrsd}) convolved with the window function using the
method described in the following sub-section.

\subsubsection{Convolution by the window function}

The expectation value of the power spectrum estimators in Section
\ref{secpkmultest} is the underlying power spectrum $P(\vec{k})$
convolved with the survey selection function.  These convolutions may
also be evaluated using FFTs, which we accomplished using the
following scheme extending the results of the previous section:
\begin{eqnarray}
& & P_{0,{\rm c}}(\vec{k}) = \frac{1}{I} (A_0 A_0^*)_{\rm c} , \\
& & P_{2,{\rm c}}(\vec{k}) = \frac{5}{2I} \left[ 3 (A_0 A_2^*)_{\rm c} - (A_0 A_0^*)_{\rm c} \right] , \\
& & P_{4,{\rm c}}(\vec{k}) = \frac{9}{8I} \left[ 35 (A_0 A_4^*)_{\rm c} - 30 (A_0 A_2^*)_{\rm c} + 3 (A_0 A_0^*)_{\rm c} \right] .
\end{eqnarray}
where
\begin{eqnarray}
& & (A_0 A_0^*)_{\rm c} = \int d^3\vec{k}' \, P_{\rm mod}(\vec{k}') \, W_0(\delta \vec{k}) \, W_0^*(\delta \vec{k}) , \\
& & (A_0 A_2^*)_{\rm c} = \int d^3\vec{k}' \, P_{\rm mod}(\vec{k}') \, W_0(\delta \vec{k}) \, \times \nonumber \\
& & \sum_{ij} \beta_{ij}(\delta \vec{k}) \, W_{2,ij}^*(\delta \vec{k}) , \\
& & (A_0 A_4^*)_{\rm c} = \int d^3\vec{k}' \, P_{\rm mod}(\vec{k}') \, W_0(\delta \vec{k}) \, \times \nonumber \\
& & \sum_{ijk} \gamma_{ijk}(\delta \vec{k}) \, W_{4,ijk}^*(\delta \vec{k}) ,
\end{eqnarray}
where $\delta \vec{k} = \vec{k} - \vec{k}'$, in terms of
\begin{eqnarray}
& & W_0(\vec{r}) = w_{\rm FKP}(\vec{r}) \, W(\vec{r}) , \\
& & W_{2,ij}(\vec{r}) = b_{ij}(\vec{r}) \, w_{\rm FKP}(\vec{r}) \, W(\vec{r}) , \\
& & W_{4,ijk}(\vec{r}) = c_{ijk}(\vec{r}) \, w_{\rm FKP}(\vec{r}) \, W(\vec{r}) .
\end{eqnarray}
For reasons of further computing speed when fitting models, we re-cast
this convolution as a matrix multiplication in Fourier bins of width
$\Delta k = 0.02 \, h$ Mpc$^{-1}$:
\begin{equation}
P_{\rm est}(i) = \sum_j M_{ij} \, P_{\rm mod}(j) ,
\label{eqpkconv}
\end{equation}
where the arrays $P_{\rm est}(i)$ and $P_{\rm mod}(j)$ consist of the
concatenation
\begin{equation}
\lbrace P_0(k_1), P_0(k_2), ..., P_2(k_1), P_2(k_2), ..., P_4(k_1),
P_4(k_2), ... \rbrace .
\label{eqpkconcat}
\end{equation}
We determined the matrix $M_{ij}$ by evaluating the full convolution
for a set of unit vectors.  For each unit vector the model $P_{\rm
  mod}(\vec{k})$ is computed using Equation \ref{eqpkmultdef1} setting
a single element of Equation \ref{eqpkconcat} to unity and the rest
of the elements to zero.

\subsubsection{Covariance matrix}

\begin{figure*}
\begin{center}
\resizebox{12cm}{!}{\rotatebox{270}{\includegraphics{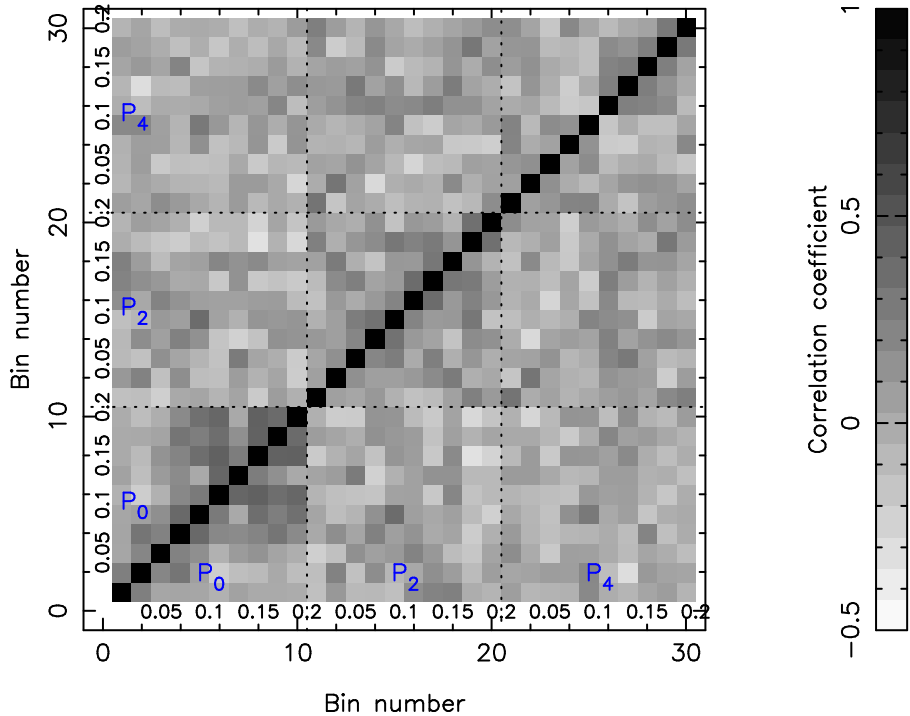}}}
\end{center}
\caption{The correlation matrix for the 2dFLenS power spectrum
  multipoles arranged in a data vector $P_{\rm est}(i) = \lbrace P_0,
  P_2, P_4 \rbrace$, derived from the mock catalogues as ${\rm
    Cov}_{ij}/\sqrt{{\rm Cov}_{ii} \, {\rm Cov}_{jj}}$.  Results are
  shown for the redshift range $0.43<z<0.7$ for the KiDS-S region;
  they are similar for the other region and redshift range.  The
  labels $(0.05, 0.1, 0.15, 0.2)$ denote wavenumbers in $h$
  Mpc$^{-1}$.}
\label{figpkmultcov}
\end{figure*}

We measured the power spectrum multipoles for each 2dFLenS mock
catalogue, producing a series of data vectors $P_{\rm est}(i)$.  We
hence deduced the covariance matrix by averaging over the mocks
\begin{equation}
{\rm Cov}_{ij} = \langle P_{\rm est}(i) \, P_{\rm est}(j) \rangle -
\langle P_{\rm est}(i) \rangle \, \langle P_{\rm est}(j) \rangle .
\end{equation}
The corresponding correlation matrix, defined by ${\rm
  Cov}_{ij}/\sqrt{{\rm Cov}_{ii} \, {\rm Cov}_{jj}}$, is displayed in
Figure \ref{figpkmultcov} for the KiDS-S region for the redshift range
$0.43 < z < 0.7$ (the other region and redshift range are similar).

\subsection{Redshift-space distortion}
\label{secrsd}

We fitted the measured 2dFLenS clustering multipoles using a standard
model for the redshift-space galaxy power spectrum as a function of
the angle of the Fourier wavevector to the line-of-sight:
\begin{equation}
P_{\rm g}(k,\mu) = b^2 \, P_{\rm m}(k) \, (1 + \beta \mu^2)^2 \,
\exp{(-k^2 \mu^2 \sigma_v^2/H_0^2)} ,
\label{eqmodrsd}
\end{equation}
where $b$ is the galaxy bias factor, $P_{\rm m}(k)$ is the fiducial
non-linear matter power spectrum defined in Section \ref{secprojcorr}
and $\beta = f/b$ parameterizes the amplitude of redshift-space
distortion in terms of the growth rate of structure $f$.\footnote{We
  prefer to fit for $\beta$ in this Section, rather than for $f$,
  because $\beta$ is required as an input for the gravitational slip
  measurements presented by Amon et al.\ (2016).}  This model combines
the large-scale `Kaiser limit' amplitude correction (Kaiser 1987) with
a heuristic damping of power on smaller scales that describes a
leading-order perturbation theory correction (Scoccimarro 2004) in
terms of a free parameter $\sigma_v$ with units of km s$^{-1}$.  Our
model is hence characterized by three parameters $(\beta, \sigma_v,
b)$.

We fitted this 3-parameter model to the monopole and quadrupole of
both the power spectrum and correlation function in each analysis
region.  For given values of $(\beta, \sigma_v, b)$ we deduced the
unconvolved model power spectrum multipoles $P_\ell(k)$ from
$P(k,\mu)$ using Equation \ref{eqpkmultdef2}, which we convolved with
the survey window function using Equation \ref{eqpkconv}.  The model
correlation function multipoles may be determined from $P_\ell(k)$
using
\begin{equation}
\xi_\ell(s) = \frac{i^\ell}{2 \pi^2} \int dk \, k^2 \, P_\ell(k) \,
j_\ell(ks) ,
\end{equation}
where $j_\ell$ is the spherical Bessel function of order $\ell$.  We
performed the fits by evaluating the $\chi^2$ statistic of each model
using the full covariance matrix.  For example, for the case of the
power spectrum multipoles we determined:
\begin{equation}
\chi^2 = \sum_{ij} \left[ P_{\rm est}(i) - P_{\rm mod}(i) \right]
\left[ {\rm Cov}^{-1} \right]_{ij} \left[ P_{\rm est}(j) - P_{\rm
  mod}(j) \right]
\end{equation}
for each analysis region, and summed the $\chi^2$ values.  We
propagated the errors induced by estimating the inverse of an $N_{\rm
  bin} \times N_{\rm bin}$ covariance matrix from a limited number of
mock catalogues $N_{\rm mock} = 65$ by computing the correction
determined by Sellentin \& Heavens (2016), in which the likelihood of
each model is given by
\begin{equation}
{\rm Likelihood} \propto \left( 1 + \frac{\chi^2}{N_{\rm mock}-1}
\right)^{-N_{\rm mock}/2} .
\end{equation}
Our analyses utilized at most $N_{\rm bin} \sim 18$ data points, such
that the number of 65 mocks was adequate.

\begin{figure}
\begin{center}
\resizebox{8cm}{!}{\rotatebox{270}{\includegraphics{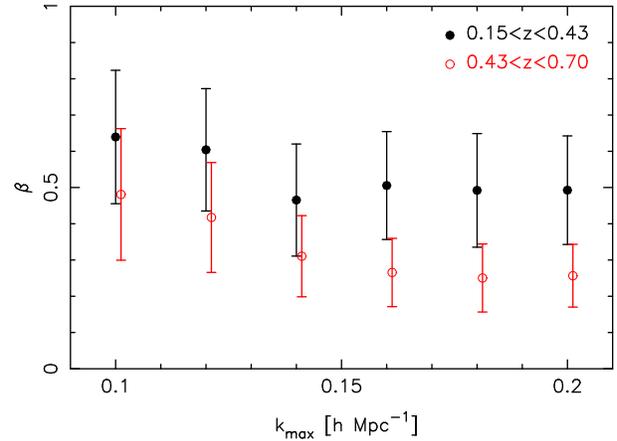}}}
\end{center}
\caption{Dependence of the marginalized measurements of $\beta$ on the
  maximum wavenumber used when fitting the 2dFLenS power spectrum
  multipole data.  Results are shown for the two redshift ranges $0.15
  < z < 0.43$ (solid black circles) and $0.43 < z < 0.70$ (open red
  circles).  Measurements for the different redshifts are slightly
  shifted along the $x$-axis for clarity.}
\label{figbetakmax}
\end{figure}

We generated our baseline model fits using the power spectrum
multipole measurements in the range $0.02 < k < 0.2 \, h$ Mpc$^{-1}$.
Our results are not particularly sensitive to the fitting range:
Figure \ref{figbetakmax} demonstrates the low sensitivity of the
best-fitting measurement of $\beta$ to the maximum wavenumber used in
the fit, with the variation being encompassed by the size of the
statistical errors.  We prefer to demonstrate the robustness of our
results in this manner, rather than by fitting to our mocks, because
the mock galaxy catalogues lack a satellite population hence may not
be reliable for this purpose.

\begin{figure}
\begin{center}
\resizebox{8cm}{!}{\rotatebox{270}{\includegraphics{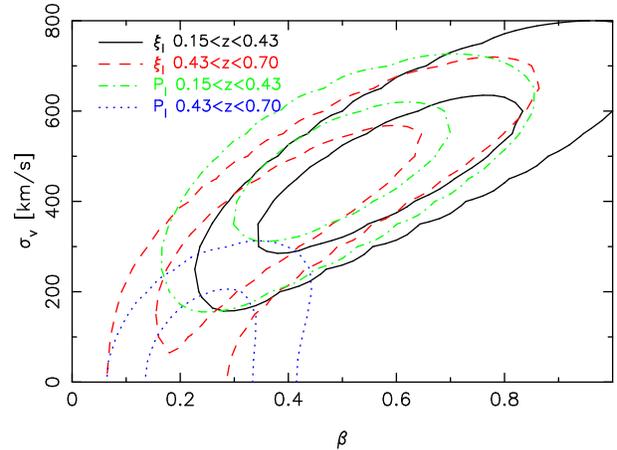}}}
\end{center}
\caption{Confidence regions of fits for $\beta$ and $\sigma_v$,
  marginalizing over $b$.  Results are shown for fits to both
  correlation function and power spectrum multipoles, for two redshift
  ranges $0.15<z<0.43$ and $0.43<z<0.70$.  $68\%$ and $95\%$
  confidence regions are shown in all cases.}
\label{figparfits}
\end{figure}

Our marginalized parameter measurements for the $0.15 < z < 0.43$
datasets are:
\begin{eqnarray}
\beta &=& 0.49 \pm 0.15 , \\
\sigma_v &=& 470 \pm 110 \; {\rm km/s} , \\
b &=& 1.9 \pm 0.1 .
\end{eqnarray}
For $0.43 < z < 0.7$ we obtain:
\begin{eqnarray}
\beta &=& 0.26 \pm 0.09 , \\
\sigma_v &=& 100 \pm 100 \; {\rm km/s} , \\
b &=& 2.2 \pm 0.1 .
\end{eqnarray}
The best-fitting models are overplotted in Figure \ref{figpkmultdata}.
The corresponding chi-squared values for the two redshift ranges are
$37.1$ and $32.6$, for 33 degrees of freedom, indicating that the
models are a good fit to the data, and the best-fitting bias
parameters are consistent with those obtained from the projected
correlation function in Section \ref{secprojcorr}.  Figure
\ref{figparfits} displays the 2D posterior probability distribution of
$(\beta,\sigma_v)$ for redshift ranges $0.15 < z < 0.43$ and $0.43 < z
< 0.7$.

We can compare our $\beta$ measurements to those determined using BOSS
LRGs: Sanchez et al.\ (2014) report $\beta_{\rm LOWZ} = 0.38 \pm 0.11$
and $\beta_{\rm CMASS} = 0.36 \pm 0.06$.  Our measurements are
consistent, albeit with a $\sim 50\%$ larger statistical error.
However, the weak lensing data overlapping these 2dFLenS measurements
permit some unique applications of our results (Joudaki et al.\ 2016b,
Amon et al.\ 2016).  The amplitudes of our measured redshift-space
distortion parameters are also consistent, given their statistical
errors, with those predicted by the growth rate in the standard
$\Lambda$CDM cosmological model and our best-fitting galaxy bias
factors.

As a point of comparison, we also fitted the RSD models to our
correlation function multipole measurements in the range $10 < s < 90
\, h^{-1}$ Mpc (again, we note that our results are not particularly
sensitive to the fitting range).  We overplot the parameter
constraints in Figure \ref{figparfits}, illustrating that the power
spectrum and correlation function multipoles produce consistent
results.  The best-fitting correlation function multipole models are
overplotted in Figures \ref{figximultdata}.

\section{Summary}
\label{secconc}

In this paper we have introduced the 2-degree Field Lensing Survey
(2dFLenS), a new galaxy redshift survey performed at the
Anglo-Australian Telescope which extends the spectroscopic-redshift
coverage of gravitational lensing surveys in the southern sky, with a
particular focus on the overlapping Kilo-Degree Survey (KiDS).
2dFLenS contains $70{,}079$ objects with good-quality redshifts,
including $40{,}531$ Luminous Red Galaxies and $28{,}269$ objects
which form a magnitude-limited nearly-complete sub-sample.  The LRGs
may be utilized for analysis of galaxy-galaxy lensing, redshift-space
distortion and determination of the imaging source redshift
distribution by cross-correlation, and the magnitude-limited sample
may be employed for direct source classification and
photometric-redshift calibration.

In this paper we have presented the survey selection function and
clustering measurements for the LRG sample and corresponding mock
catalogues.  We fitted redshift-space distortion models to the
clustering multipoles, finding RSD parameters $\beta = 0.49 \pm 0.15$
and $0.26 \pm 0.09$ for redshift ranges $0.15 < z < 0.43$ and $0.43 <
z < 0.7$, respectively.  These values are consistent with those
obtained from LRGs in the Baryon Oscillation Spectroscopic Survey, and
(when combined with the best-fitting galaxy bias factors), consistent
with the predictions of the standard $\Lambda$CDM cosmological model.

Five associate science papers are currently in preparation:
\begin{itemize}
\item {\it Johnson et al.\ (2016)} present a new quadratic-estimation
  method for constraining the source redshift distribution of an
  imaging survey via cross-correlations with a spectroscopic redshift
  survey, with an application to KiDS and 2dFLenS data.
\item {\it Joudaki et al.\ (2016b)} perform self-consistent
  cosmological model fits to overlapping cosmic shear, galaxy-galaxy
  lensing and redshift-space distortion data from KiDS and 2dFLenS.
\item {\it Amon et al.\ (2016)} determine new measurements of the
  gravitational slip statistic, $E_G$, to large scales, using data
  from KiDS and 2dFLenS.
\item {\it Wolf et al.\ (2016)} use the magnitude-limited sample of
  2dFLenS redshifts to compare various techniques for direct
  photometric-redshift calibration based on kernel-density estimation,
  machine learning with neural networks, and template fits.
\item {\it Janssens et al.\ (2016)} analyze the `red-nugget'
  spare-fibre sample to place new constraints on the redshift
  evolution of compact early-type galaxies.
\end{itemize}
2dFLenS data products will be released with the publication of these
papers via our website {\tt http://2dflens.swin.edu.au}.

\section*{Acknowledgments}

This research made use of Astropy, a community-developed core Python
package for Astronomy (Astropy Collaboration, 2013).

This work makes use of the {\sc runz} redshifting code developed by
Will Sutherland, Will Saunders, Russell Cannon and Scott Croom.

CB acknowledges the support of the Australian Research Council through
the award of a Future Fellowship.

JHD acknowledges support from the European Commission under a
Marie-Sk{\l}odwoska-Curie European Fellowship (EU project 656869) and
from the NSERC of Canada.

CH acknowledges funding from the European Research Council under grant
number 647112.

HH is supported by an Emmy Noether grant (No.\ Hi 1495/2-1) of the
Deutsche Forschungsgemeinschaft.

SJa acknowledges support from the Natural Sciences and Engineering
Research Council of Canada, and the Government of Ontario.

KK is supported by a grant from the Netherlands Foundation for 
Scientific Research (NWO).

DP is supported by an Australian Research Council Future Fellowship
[grant number FT130101086].

We also acknowledge the Aspen Center for Physics (NSF grant 1066293)
where some of the target selection work took place.

The 2dFLenS survey is based on data acquired through the Australian
Astronomical Observatory, under program A/2014B/008.  It would not
have been possible without the dedicated work of the staff of the AAO
in the development and support of the 2dF-AAOmega system, and the
running of the AAT.

Computations for the $N$-body simulations were performed in part on
the Orcinus supercomputer at the WestGrid HPC consortium ({\tt
  www.westgrid.ca}), in part on the GPC supercomputer at the SciNet
HPC Consortium. SciNet is funded by: the Canada Foundation for
Innovation under the auspices of Compute Canada; the Government of
Ontario; Ontario Research Fund - Research Excellence; and the
University of Toronto.

{\it Author contributions:} All authors contributed to the development
and writing of this paper.  The authorship list reflects the lead
author of this paper (CB) followed by an alphabetical group.  This
group includes key contributors to the target selection (TE, CH, HH,
DK, CW), AAT observing (AA, MC, KG, AJ, SJa, SJo, CL, FM, DP),
redshifting (AA, MC, TE, KG, SH, AJ, SJo, KK, CL, FM, DP, GP, CW) and
N-body simulations (JHD).

\appendix

\section{Magnitude transformations}
\label{secmagtrans}

The 2dFLenS LRG selection criteria are inspired by the SDSS, BOSS and
eBOSS surveys.  Since the filter systems used by the optical imaging
surveys used to select 2dFLenS targets -- VST OmegaCAM and CFHT
MegaCam -- are not identical to SDSS filters, we derived magnitude
transformations between the different filter systems using an
elliptical galaxy template spectrum.

We refer to ATLAS magnitudes as $(u_A, g_A, r_A, i_A, z_A)$, CFHT
magnitudes as $(u_C, g_C, r_C, i_C, z_C)$ or $(u_C, g_C, r_C, y_C,
z_C)$ (depending on which version of the $i$-band filter was used for
a pointing, as described by Erben et al.\ 2013) and SDSS magnitudes as
$(u_S, g_S, r_S, i_S, z_S)$.  Our template spectrum then yielded
transformations
{\tiny
\[
\left( \begin{array}{c}
u_S \\
g_S \\
r_S \\
i_S \\
z_S
\end{array} \right)
=
\left( \begin{array}{ccccc}
1.0121 & -0.0123 & 0.0001 & 0 & 0 \\
0 & 1.0091 & -0.0103 & 0.0012 & 0 \\
0 & 0 & 1.1297 & -0.1297 & 0 \\
0 & 0 & 0.0308 & 0.9692 & 0 \\
0 & 0 & -0.0008 & -0.0249 & 1.0256
\end{array} \right)
\left( \begin{array}{c}
u_A \\
g_A \\
r_A \\
i_A \\
z_A
\end{array} \right)
\]
\[
\left( \begin{array}{c}
u_S \\
g_S \\
r_S \\
i_S \\
z_S
\end{array} \right)
=
\left( \begin{array}{ccccc}
1.2674 & -0.3095 & 0.0442 & -0.0021 & 0 \\
0 & 1.1574 & -0.1651 & 0.0077 & 0 \\
0 & 0 & 1.0491 & -0.0491 & 0 \\
0 & 0 & 0.1057 & 0.8943 & 0 \\
0 & 0 & -0.0087 & -0.0736 & 1.0823
\end{array} \right)
\left( \begin{array}{c}
u_C \\
g_C \\
r_C \\
i_C \\
z_C
\end{array} \right)
\]
\[
\left( \begin{array}{c}
u_S \\
g_S \\
r_S \\
i_S \\
z_S
\end{array} \right)
=
\left( \begin{array}{ccccc}
1.2674 & -0.3095 & 0.0443 & -0.0022 & 0 \\
0 & 1.1574 & -0.1656 & 0.0082 & 0 \\
0 & 0 & 1.0520 & -0.0520 & 0 \\
0 & 0 & 0.0520 & 0.9480 & 0 \\
0 & 0 & -0.0043 & -0.0780 & 1.0823
\end{array} \right)
\left( \begin{array}{c}
u_C \\
g_C \\
r_C \\
y_C \\
z_C
\end{array} \right)
\]
}
The colour coefficients have very little variation with redshift; we
use average values in the redshift range $0.15 < z < 0.7$.  Using
these transformation matrices we obtained the following relations for
the LRG colour variables defined by Equations \ref{eqcpar},
\ref{eqcperp} and \ref{eqdperp}:
\begin{eqnarray}
c_\parallel &=& 0.7064 \, g_A + 0.5207 \, r_A - 1.2271 \, i_A - 0.216 \nonumber \\
&=& 0.8102 \, g_C + 0.2821 \, r_C - 1.0923 \, i_C - 0.216 \nonumber \\
&=& 0.8102 \, g_C + 0.3477 \, r_C - 1.1579 \, y_C - 0.216 ,
\end{eqnarray}
\begin{eqnarray}
c_\perp &=& -0.2523 \, g_A + 1.3839 \, r_A - 1.1316 \, i_A - 0.18 \nonumber \\
&=& -0.2894 \, g_C + 1.2469 \, r_C - 0.9576 \, i_C - 0.18 \nonumber \\
&=& -0.2894 \, g_C + 1.3044 \, r_C - 1.0150 \, y_C - 0.18 ,
\end{eqnarray}
\begin{eqnarray}
d_\perp &=& -0.1261 \, g_A + 1.2414 \, r_A - 1.1153 \, i_A \nonumber \\
&=& -0.1447 \, g_C + 1.0952 \, r_C - 0.9505 \, i_C \nonumber \\
&=& -0.1447 \, g_C + 1.1522 \, r_C - 1.0075 \, y_C .
\end{eqnarray}


\begin{thebibliography}{}
\bibitem{1} Ahn, C., et al., 2014, ApJS, 211, 17
\bibitem{2} Alam, S., et al., 2016, submitted
\bibitem{3} Amon, A., et al., 2016, in preparation
\bibitem{4} Anderson, L., et al., 2014, MNRAS, 441, 24
\bibitem{5} Aubourg, E., et al., 2015, PhRvD, 92, 3516
\bibitem{6} Baldry, I., et al., 2014, MNRAS, 441, 2440
\bibitem{7} Becker, M., et al., 2016, PhRvD, 94, 2002
\bibitem{8} Bertin, E., Arnouts, S., 1996, AS, 117, 393
\bibitem{9} Betoule, M., et al., 2014, A\&A, 568, 22
\bibitem{10} Beutler, F., et al., 2011, MNRAS, 416, 3017
\bibitem{11} Beutler, F., et al., 2012, MNRAS, 423, 3430
\bibitem{12} Beutler, F., et al., 2014, MNRAS, 443, 1065
\bibitem{13} Bianchi, D., Gil-Marin, H., Ruggeri, R., Percival, W.,
  2015, MNRAS, 453, 11
\bibitem{14} Blake, C., Wall, J., 2002, MNRAS, 337, 993
\bibitem{15} Blake, C., et al., 2011a, MNRAS, 415, 2876
\bibitem{16} Blake, C., et al., 2011b, MNRAS, 418, 1707
\bibitem{17} Blake, C., et al., 2016, MNRAS, 456, 2806
\bibitem{18} Bleem, L., et al., 2015, ApJS, 216, 27
\bibitem{19} Buddendiek, A., et al., 2016, MNRAS, 456, 3886
\bibitem{20} Cai, Y.-C., Bernstein, G., 2012, MNRAS, 422, 1045
\bibitem{21} Cannon, R., et al., 2006, MNRAS, 372, 425
\bibitem{22} Choi, A., et al., 2016, MNRAS, submitted
\bibitem{23} Chow, N., Khoury, J., PhRvD, 80, 4037
\bibitem{24} Colless, M., et al., 2001, MNRAS, 328, 1039
\bibitem{25} Conley, A., et al., 2011, ApJS, 192, 1
\bibitem{26} Cunha, C., Huterer, D., Busha, M., Wechsler, R., 2012,
  MNRAS, 423, 909
\bibitem{27} Damjanov, I., et al., 2009, ApJ, 695, 101
\bibitem{28} Damjanov, I., et al., 2015, ApJ, 806, 158
\bibitem{29} Dawson, K., et al., 2013, AJ, 145, 10
\bibitem{30} Dawson, K., et al., 2016, AJ, 151, 44
\bibitem{31} de Jong, J., et al., 2015, A\&A, 582, 62
\bibitem{32} de la Torre, S., et al., 2013, A\&A, 557, 54
\bibitem{33} Delubac, T., et al., 2015, A\&A, 574, 59
\bibitem{34} de Putter, R., Dore, O., Das, S., 2014, ApJ, 780, 185
\bibitem{35} de Rham, C., et al., 2008, PhRvL, 100, 1603
\bibitem{36} Drinkwater, M., et al., 2010, MNRAS, 401, 1429
\bibitem{37} Driver, S., et al., 2011, MNRAS, 413, 971
\bibitem{38} Eisenstein, D., et al., 2001, AJ, 122, 2267
\bibitem{39} Erben, T., et al., 2005, AN, 326, 432
\bibitem{40} Erben, T., et al., 2009, A\&A, 493, 1197
\bibitem{41} Erben, T., et al., 2013, MNRAS, 433, 2545
\bibitem{42} Eriksen, M., Gaztanaga, E., 2015, MNRAS, 451, 1553
\bibitem{43} Feldman, H., Kaiser, N., Peacock, J., 1994, ApJ, 426,
  23
\bibitem{44} Fixsen, D., Cheng, E., Gales, J., Mather, J., Shafer, R.,
  Wright, E., 1996, ApJ, 473, 576
\bibitem{45} Font-Ribera, A., McDonald, P., Mostek, N., Reid, B., Seo,
  H.-J., Slosar, A., 2014, JCAP, 5, 23
\bibitem{46} Gaztanaga, E., Eriksen, M., Crocce, M., Castander, F.,
  Fosalba, P., Marti, P., Miquel, R., Cabre, A., 2012, MNRAS, 422,
  2904
\bibitem{47} Gilbank, D., Gladders, M., Yee, H., Hsieh, B., 2011, AJ,
  141, 94
\bibitem{48} Guo, H., et al., 2014, MNRAS, 441, 2398
\bibitem{49} Gwyn, S., 2012, AJ, 143, 38
\bibitem{50} Harnois-Deraps, J., Pen, U.-L., Iliev, I., Merz, H.,
  Emberson, J., Desjacques, V., 2013, MNRAS, 436, 540
\bibitem{51} Harnois-Deraps, J., van Waerbeke, L., 2015, MNRAS, 450,
  2857
\bibitem{52} Heymans, C., et al., 2012, MNRAS, 427, 146
\bibitem{53} Hildebrandt, H., et al., 2012, MNRAS, 421, 2355
\bibitem{54} Hildebrandt, H., et al., 2016a, MNRAS, submitted
\bibitem{55} Hildebrandt, H., et al., 2016b, MNRAS, submitted
\bibitem{56} Hinton, S., Davis, T., Lidman, C., Glazebrook, K., Lewis,
  G., 2016, Astronomy \& Computing, 15, 61
\bibitem{57} Huff, E., Eifler, T., Hirata, C., Mandelbaum, R.,
  Schlegel, D., Seljack, U., 2014, MNRAS, 440, 1322
\bibitem{58} Janssens, S., et al., 2016, in preparation
\bibitem{59} Johnson, A., et al., 2016, in preparation
\bibitem{60} Joudaki, S., et al., 2016a, submitted
\bibitem{61} Joudaki, S., et al., 2016b, in preparation
\bibitem{62} Kaiser, N., 1987, MNRAS, 227, 1
\bibitem{63} Kaiser, N., Squires, G., Broadhurst, T., 1995, ApJ, 449, 460
\bibitem{64} Kazin, E., et al., 2014, MNRAS, 441, 3524
\bibitem{65} Kirk, D., Lahav, O., Bridle, S., Jouvel, S., Abdalla, F.,
  Frieman, J., 2015, MNRAS, 451, 4424
\bibitem{66} Kuijken, K., et al., 2015, MNRAS, 454, 3500
\bibitem{67} Landy S., Szalay A., 1993, ApJ, 412, 64
\bibitem{68} Lewis, A., Challinor, A., Lasenby, A., 2000, ApJ, 538,
  473
\bibitem{69} Lewis, I., et al., 2002, MNRAS, 333, 279
\bibitem{70} Lidman, C., et al., 2016, in preparation
\bibitem{71} Ma, Z., Hu, W., Huterer, D., 2006, ApJ, 636, 21
\bibitem{72} Mandelbaum, R., et al., 2013, MNRAS, 432, 1544
\bibitem{73} Marin, F., Beutler, F., Blake, C., Koda, J., Kazin, E.,
  Schneider, D., 2016, MNRAS, 455, 4046
\bibitem{74} McDonald, P., Seljak, U., 2009, JCAP, 10, 7
\bibitem{75} McQuinn, M., White, M., 2013, MNRAS, 433, 2857
\bibitem{76} Mehrtens, N., et al., 2012, MNRAS, 423, 1024
\bibitem{77} Miszalski, B., et al., 2006, MNRAS, 371, 1537
\bibitem{78} More, S., Miyatake, H., Mandelbaum, R., Takada, M.,
  Spergel, D., Brownstein, J., Schneider, D., 2015, ApJ, 806, 2
\bibitem{79} Newman, J., et al., 2015, APh, 63, 81
\bibitem{80} Pickles, A., 1998, PASP, 110, 863
\bibitem{81} {\sl Planck} collaboration (2015 results XIII), submitted
\bibitem{82} Poggianti, B., et al., 2013, ApJ, 762, 77
\bibitem{83} Prakash, A., Licquia, T., Newman, J., Rao, S., 2015, ApJ,
  803, 105
\bibitem{84} Reyes, R., et al., 2010, Nature, 464, 256
\bibitem{85} Samushia, L., et al., 2014, MNRAS, 439, 3504
\bibitem{86} Sanchez, A., et al., 2014, MNRAS, 440, 2692
\bibitem{87} Saunders, W., et al., 2004, SPIE, 5492, 389
\bibitem{88} Saunders, W., Cannon, R., Sutherland, W., 2004,
  Anglo-Australian Observatory Newsletter, 106, 16
\bibitem{89} Schirmer, M., 2013, ApJS, 209, 21
\bibitem{90} Schlegel, D., Finkbeiner, D., Davis, M., 1998, ApJ, 500,
  525
\bibitem{91} Scoccimarro, R., 2004, PhRvD, 70, 83007
\bibitem{92} Scoccimarro, R., 2015, PhRvD, 92, 3532
\bibitem{93} Sellentin, E., Heavens, A., 2016, MNRAS, 456, 132
\bibitem{94} Shanks, T., et al., 2015, MNRAS, 451, 4238
\bibitem{95} Sharp et al., 2006, SPIE, 6269, 14
\bibitem{96} Simpson, F., et al., 2013, MNRAS, 429, 2249
\bibitem{97} Skrutskie, M., et al., 2006, AJ, 131, 1163
\bibitem{98} Song, Y.-S., et al., 2011, PhRvD, 84, 3523
\bibitem{99} Sotiriou, T., Faraoni, V., 2010, RvMP, 82, 451
\bibitem{100} Suzuki, N., et al., 2012, ApJ, 746, 85
\bibitem{101} Takahashi, R., Sato, M., Nishimichi, T., Taruya, A.,
  Oguri, M., 2012, ApJ, 761, 152
\bibitem{102} Taylor, E., et al., 2010, ApJ, 720, 723
\bibitem{103} Tonry, J., Davis, M., 1979, AJ, 84, 1511
\bibitem{104} Valentinuzzi, T., et al., 2010, ApJ, 712, 226
\bibitem{105} van Dokkum, P., et al., 2008, ApJ, 677, 5
\bibitem{106} Wolf, C., et al., 2016, in preparation
\bibitem{107} Yamamoto, K., Nakamichi, M., Kamino, A., Bassett, B.,
  Nishioka, H., 2006, PASJ, 58, 93
\bibitem{108} Yanny, B., et al., 2009, ApJ, 700, 1282
\bibitem{109} Yuan, F., et al., 2015, MNRAS, 452, 3047
\bibitem{110} Zacharias, N., Finch, C., Girard, T., Henden, A.,
  Bartlett, J., Monet, D., Zacharias, M., 2013, AJ, 145, 44
\bibitem{111} Zhang, P., et al., 2007, PhRvL, 99, 1302
\end{thebibliography}
\end{document}